\documentclass[useAMS,usenatbib]{mnras}
\usepackage[pdftex]{graphicx} 
\usepackage{hyperref}
\usepackage{xcolor} 
\usepackage{lscape}
\usepackage{calc}
\usepackage{amsmath}
\usepackage{afterpage}
\usepackage{bm}
\usepackage{amssymb, physics}

\hypersetup{colorlinks,linkcolor={red!50!black}, citecolor={blue!50!black},urlcolor={blue!80!black} } 

\title[Drivers of galaxy quenching]
{In-situ vs. ex-situ drivers of galaxy quenching: critical black hole mass and main sequence universality in the FLAMINGO simulation}

\author[Lim et al.]
{Seunghwan Lim$^{1,2}$\thanks{E-mail: sl2207@cam.ac.uk},
Sandro Tacchella$^{1,2}$, 
Roberto Maiolino$^{1,2}$, 
Joop Schaye$^{3}$, 
Matthieu Schaller$^{4,3}$ 
\\
\vspace*{6pt} \\
$^{1}$Kavli Institute for Cosmology, University of Cambridge, Madingley
Road, Cambridge, CB3 0HA, UK \\
$^{2}$Cavendish Laboratory, University of Cambridge, 19 JJ Thomson
Avenue, Cambridge, CB3 0HE, UK \\
$^{3}$Leiden Observatory, Leiden University, PO Box 9513, 2300 RA Leiden, the Netherlands \\
$^{4}$Lorentz Institute for Theoretical Physics, Leiden University, PO box 9506, 2300 RA Leiden, the Netherlands 
\
}

\begin{document} 

\pagerange{\pageref{firstpage}--\pageref{lastpage}}

\date{\today}
\pubyear{2025}

\maketitle

\label{firstpage}

\begin{abstract}
Exploiting a large sample of 5.3 million galaxies with $M_\ast\,{=}\,10^{10-11}\,{\rm M}_\odot$ from the highest-resolution FLAMINGO simulation, we carry out a statistical analysis of quiescent and star-forming galaxies to explore quenching mechanisms. From redshift $z\,{\simeq}\,7$ to 0, we find that the median star-formation rate of main-sequence galaxies is independent of the environment and of whether a galaxy is a central or satellite, whereas the fraction of quiescent galaxies is highly sensitive to both. By employing Random Forest (RF) classifiers, we demonstrate that black hole (BH) feedback is the most responsible quenching mechanism for both centrals and satellites, while halo mass is the second most significant. For satellites, a notable importance given by RF to stellar mass implies in-situ pre-quenching rather than ex-situ preprocessing prior to infall to the current host halo. In the cosmic afternoon of $z\,{=}\,$0\,--\,1, we identify two distinct regimes of evolution: at $M_{\rm BH}\,{\gtrsim}\,10^7\,{\rm M}_\odot$, essentially all galaxies are quenched regardless of their environment; at $M_{\rm BH}\,{\lesssim}\,10^7\,{\rm M}_\odot$, quenching is determined mainly by halo mass. Galaxies undergo a sharp transition from the main sequence to quiescence once their BH mass reaches $M_{\rm BH}\,{\simeq}\,10^7\,{\rm M}_\odot$ (typically when $M_\ast\,{\simeq}\,10^{10.5}\,{\rm M}_\odot$ and $M_{\rm h}\,{\simeq}\,10^{12}\,{\rm M}_\odot$) with a short quenching timescale of ${<}$\,1\,Gyr. This transition is driven by a sudden change in the gas mass in the inner circum-galactic medium. Our results indicate that galaxy quenching arises from a combination of in-situ and ex-situ physical processes. 
\end{abstract} 

\begin{keywords}
methods: statistical -- galaxies: formation -- galaxies: evolution -- galaxies: clusters: general -- galaxies: haloes
\end{keywords}

\section[intro]{Introduction}
\label{sec_intro}

In contemporary theories of galaxy evolution, star and galaxy formation are driven by gas falling into dark matter structures, which have collapsed from the seeds of primordial density fluctuations \citep[see][for a review]{MovdBWhite2010}. This process is followed by cooling through radiation. Early models based on this simplified framework, however, failed to reproduce realistic galaxies that align with observations \citep[e.g.][]{WhiteRees1978, DekelSilk1986, Cole1994, MacLowFerrara1999}. Even Milky Way-mass galaxies—the most prevalent type in the Universe—were not accurately modeled, with simulations producing an excessive number of stars and overly massive galaxies \citep[e.g.,][]{Balogh2001, Dave2001}. This issue, known as the overcooling problem \citep{WhiteRees1978, WhiteFrenk1991}, was later addressed gradually in refined theoretical models \citep[e.g.][]{Bower2006, Croton2006, Sijacki2007, McCarthy2010, GaborDave2012, Vogelsberger2014, Schaye2015} that incorporate substantial energy input from supernova explosions and stellar winds \citep[e.g.,][]{Larson1974, DekelSilk1986, SpringelHernquist2003, Veilleux2005}, and active galactic nucleus (AGN) \citep[e.g.,][]{BestHeckman2012, Fabian2012, HeckmanBest2014, KingPounds2015}. It is now widely accepted that such feedback mechanisms are crucial for galaxy evolution, acting as self-regulating processes that suppress star formation \citep[see][for a review]{SomervilleDave2015}. 

Star-formation efficiency, defined as the amount of stars formed relative to either the expected baryons (from the cosmic fraction) or the available gas within galaxies, has been found to be much lower than anticipated \citep[e.g.,][]{FukugitaPeebles2004}. This is in stark contrast to the scenario where most of the baryons in the cosmic fraction would have been converted into stars by the present epoch \citep[e.g.,][]{Cole2000, Bower2008}. In reality, galaxy formation efficiency---quantified as the total mass of formed stars divided by the expected baryonic mass (according to the cosmic fraction)---peaks at around 20 per cent, near the halo mass of Milky Way-size systems with $M_{\rm h}\,{\simeq}\,10^{12}\,{\rm M}_\odot$ \citep[e.g.,][]{Moster2010, Behroozi2013}. Furthermore, this efficiency decreases even further as one moves away from this mass, in either direction.

An enormous amount of energy injected by supermassive black holes (SMBHs) through AGN feedback is widely recognized as a key factor in reducing the baryon conversion efficiency in massive galaxies and halos \citep[e.g.,][]{SilkRees1998, Binney2004, ScannapiecoOh2004}. This AGN feedback is believed to suppress star formation by driving molecular and ionized outflows at a few thousands km\, s$^{-1}$ (e.g., \citealt{Sturm2011, GreeneZakamskaSmith2012, Maiolino2012, Liu2013, Perna2017, Kubo2022}), or by launching bipolar jets at relativistic speeds which carry a large amount of kinectic and thermal energy to suppress cold accretion (e.g., \citealt{McNamaraNulsen2007, Fabian2012, HeckmanBest2014, GaborDave2015, PengMaiolinoCochrane2015, BlandfordMeierReadhead2019}). Heating by X-ray radiation from the accretion disc is also thought to contribute to quenching galaxies \citep[e.g.,][]{SazonovOstrikerSunyaev2004, Hambrick2011, Choi2012}. Furthermore, accumulating theoretical and observational evidence suggests that AGN feedback has a long-term impact on galaxy evolution. Notably, \citet{Bluck2020a, Bluck2020b, Bluck2022, Bluck2023} have shown that the mass of the SMBH, an integrated quantity (or close proxies such as velocity dispersion and bulge mass), is a much better indicator of whether a galaxy is quiescent (i.e., no longer forming stars over a timescale of approximately 100 million years or more) or still actively star-forming, compared to its instantaneous properties \citep[see also e.g.,][]{Terrazas2016, Terrazas2017, MartinNavarro2018, Davies2019, Piotrowska2022}.

What complicates this relatively straightforward picture, however, is the realization that the environment of galaxies also plays a significant role in shaping their properties, including star-formation rate (SFR), color, age, morphology, and galaxy type \citep[e.g.,][]{Dressler1980, Balogh2004, Kauffmann2004, Peng2010, Lim2016, Hasan2023}. Several physical processes have been proposed to explain environmental quenching \citep[e.g.,][]{BoselliGavazzi2006}, including the sudden removal of gas via ram pressure as galaxies move through the intracluster medium (ICM) \citep[e.g.,][]{GunnGott1972}, tidal stripping of cold gas due to the gravitational influence of the halo \citep[e.g.,][]{Merritt1984}, ``strangulation" where star formation continues with currently existing gas but eventually is halted without further supply of circumgalactic medium \citep[e.g.,][]{LarsonTinsleyCaldwell1980,BaloghMorris2000,DekelBirnboim2006,PengMaiolinoCochrane2015,Baker2024}, and ``harassment" through frequent interactions with flying-by galaxies \citep[e.g.,][]{GallagherOstriker1972, Moore1996}. Disentangling the effects of environmental (or ex-situ) quenching from AGN-driven in-situ ``mass quenching" is a challenging task \citep[e.g.,][]{Balogh2016,Darvish2016,Kawinwanichakij2017,Lin2017,PintosCastro2019}, due to the vast parameter space to explore and the need for a large, unbiased sample with reliable measurements. Moreover, most of this analysis has been conducted on local galaxies at relatively low redshifts (up to $z\,{\simeq}\,0.5$--1), whereas quenching is believed to have occurred throughout much of the cosmic afternoon, at $z\,{\lesssim}\,2$. This introduces a degeneracy problem, where different models of feedback and galaxy evolution may yield similar observable signatures and predictions, thereby reducing the constraining power of observational data. The specifics of the quenching mechanisms, such as the characteristic mass and timescales over which they operate, as well as their relative strengths, may also evolve over cosmic history. Indeed, studies have reported conflicting results at $z\,{\gtrsim}\,1$ regarding the evolution of environmental quenching. Some find that, as at lower redshifts, galaxies in denser environments are redder, older, more quenched, and elliptical \citep[e.g.,][]{Patel2009}, while others argue there is no environmental dependence \citep[e.g.,][]{Scoville2013, Darvish2016} or even a reversed trend \citep[e.g.,][]{Elbaz2007, Cooper2008, Shimakawa2018}. Furthermore, significant uncertainties remain in understanding the exact mechanisms through which AGN feedback operates and how SMBHs co-evolve with their host galaxies \citep[see][for a review]{NaabOstriker2017, Werner2019}. This is underscored by recent findings from JWST, which suggest that SMBHs in the early Universe are more massive and abundant than current models predict by an order of magnitude or more \citep[e.g.,][]{Harikane2023, Ubler2023, Maiolino2024a, Maiolino2024b, Matthee2024, Scholtz2024, Tacchella2024}, with excessive number densities of massive quiescent galaxies at high redshifts \citep[e.g.,][]{Carnall2023, Valentino2023, Lagos2025, Baker2025}

In this study, we investigate a range of galaxy properties and environmental proxies and their impact on galaxy evolution, leveraging a comprehensive sample of MW-mass galaxies derived from the large-volume cosmological hydrodynamic simulation of FLAMINGO \citep{Schaye2023, Kugel2023}. Our exploration focuses on theoretical predictions to identify the dominant mechanisms and characteristic scales of galaxy evolution and quenching, with particular emphasis on the ``cosmic afternoon" at $z\,{\lesssim}2$. Specifically, we utilize Random Forest (RF) classifiers to distinguish between physical properties that directly influence galaxy quenching and those that are only correlated second-hand, while assessing their relative importance. RF has demonstrated remarkable efficacy in this context, successfully identifying direct correlations even in the presence of highly correlated secondary variables \citep[e.g.,][]{Bluck2020a, Bluck2020b, Bluck2022, Bluck2023}.

The paper is structured as follows. In Sect.~\ref{sec_data}, we introduce the FLAMINGO simulation, with a particular focus on the model features relevant to our analysis. Section.~\ref{sec_methods} describes the definition of our sample and the set-up of the RF classifiers. The primary findings are presented in Sect.~\ref{sec_result}, followed by a detailed discussion in Sect.~\ref{sec_discussion}. Finally, we summarize the results and offer conclusions in Sect.~\ref{sec_summary}.

\section[data]{The FLAMINGO simulations}
\label{sec_data}

In this work, we use the data from the Full-hydro Large-scale structure simulations with All-sky Mapping for the Interpretation of Next Generation Observations \citep[FLAMINGO;][]{Schaye2023, Kugel2023}. The FLAMINGO simulation suites are one of the largest simulations up to date, with its highest-resolution fiducial model, `L1\_m8', modeling a comoving box of $1\,{\rm cGpc}^3$ with the initial baryonic particle mass of $m_{\rm gas}=1.34\times 10^8\,{\rm M}_\odot$. As can be seen in later sections, this ensures large samples in each bin, as we explore a multi-dimensional large parameter space throughout the analysis. Although the FLAMINGO simulation suites have other runs with larger box sizes, the resolution of such boxes are only $10^9\,{\rm M}_\odot$ at best for the baryonic elements, which does not allow probing sufficiently large dynamic range and evolution for our study. We present our results based on the fiducial L1\_m8. 

FLAMINGO was run using a novel numerical solver \textsc{Swift} \citep{Schaller2024}, which uses smoothed-particle hydrodynamics (SPH) for solving hydrodynamics. The radiative cooling and heating, and the multi-phase ISM are modelled following \citet{PloeckingerSchaye2020}, and \citet{SchayeDallaVecchia2008}, respectively. For the implementation of star formation, gas particles are converted to star particles following \cite{SchayeDallaVecchia2008}, which results in matching the Kennicutt-Schmidt law. Stellar feedback such as winds and supernova (SN) is implemented via the kinetic models of \cite{Chaikin2023}. The simulations assume a \cite{Chabrier2003} IMF and the cosmology from the Dark Energy Survey year three results \citep{Abbott2022}. 

Regarding one of our main findings as will be seen later, as well as evidenced in some previous works in the literature, one of the most significant drivers for quenching galaxies is supermassive black holes (BHs) \citep[e.g.,][]{SilkRees1998, Croton2006, Somerville2008, Terrazas2020, Bluck2022}. Therefore, it is important to understand exactly how the BHs and AGN feedback thereof are implemented in the simulation. In FLAMINGO, BHs are seeded with the initial subgrid mass of $10^5\,{\rm M}_\odot$ in every halo with the total mass greater than $2.757\times 10^{11}\,{\rm M}_\odot\,(m_{\rm gas}/1.07\times10^9\,{\rm M}_\odot)$ that was not seeded in the previous snapshots, by replacing the densest gas particle still with the same dynamical (gravitational) mass at the same position. Once seeded, the BHs grow by accreting at a modified Bondi-Hoyle rate but below the Eddington limit, following \cite{Springel2005} while also boosted in high-density regions as in \cite{BoothSchaye2009} to compensate that the BH and gas are not resolved. Once a BH's subgrid mass becomes greater than its dynamical mass, the simulation follows \cite{Bahe2022} to transfer the mass from the neighboring gas particles to the BH. The BHs are repositioned manually at each time step, to the position of the gas particle with the minimum potential within three times the gravitational softening length. This accounts for the fact that the dynamical friction to hold the BHs to the gravitational center of galaxy is not modelled properly due to the numerical resolution. As a result, without repositioning, BHs tend to `float' to the outskirts of haloes in cosmological simulations, which weakens critically the impact of AGN feedback on the host galaxy's evolution \citep{Bahe2022}. Finally, following \cite{Bahe2022}, the BHs are also merged if they are within three gravitational softening lengths and below an upper limit of relative velocity. 

Whereas FLAMINGO implements two types of AGN feedback, thermal dump and jet-like kinetic model for its intermediate-resolution runs, the thermal feedback following \cite{BoothSchaye2009} is adopted as the fiducial and only model for L1\_m8. In the thermal model, the feedback energy is injected in the nearest gas particle when the accumulated amount of the energy over time steps becomes sufficient to heat the gas particle by $\Delta T_{\rm AGN}$. $\Delta T_{\rm AGN}$, the increase in the temperature of the nearest gas particle, is a model parameter that is calibrated. $\Delta T_{\rm AGN}\,{=}\,10^{8.07}\,{\rm K}$ for L1\_m8. 

The parameters of the subgrid models, including that for the AGN feedback, are tuned through a machine learning to reproduce the stellar mass function and cluster gas fraction from observations of the low-redshift Universe \citep{Kugel2023}. 

The haloes and subhaloes are identified by \textsc{VELOCIraptor} finder \citep[VR;][]{Elahi2019}. VR defines haloes by running a 3-D FoF algorithm with a linking ratio of 0.2, while subhaloes are further identified from the halo particles via FoF search in the 6-D phase space. Centrals are defined as the most distinct structure found within each FoF halo, while the other subhaloes identified by VR are considered as satellites. Unless stated otherwise, we base our analysis on the galaxy and gas properties integrated within 50\,pkpc from the center of galaxy, derived using the Spherical Overdensity and Aperture Processor (SOAP), a tool developed for the FLAMINGO project. For the halo properties, the quantities computed within $R_{\rm h}\,{=}\,R_{200}$, the 3-D radius within which the mean density is 200 times the critical density, are used. However, we confirmed that the conclusions are insensitive to the choice of aperture for both the galaxy and halo properties. For our analysis, we also use the merger tree provided as part of the FLAMINGO data product, which was constructed using the algorithm of \cite{Jiang2014} to connect the haloes between the snapshots by following the most bound particles to identify descendants and progenitors. We exploit the merger tree later for tracking the main progenitor (defined as one with the greatest number of bound particles of a descendant halo from the later snapshot) of low-redshift
haloes in question, in particular.

\section[methods]{Methods}
\label{sec_methods}

\subsection{Sample construction}
\label{ssec_sample}

For each part of our analysis, we construct samples from FLAMINGO slightly differently. This is sometimes because of the purpose of presentation, or because each part of the analysis requires a different set of predicted properties that are given zero, undefined values, or values outside the range of our interest for a subset of the whole sample. Here we describe the different types of sample construction which will be used for our analysis. The criteria and number of galaxies for the samples are summarized in Table~\ref{tab_sample}. 

\begin{figure*}
\includegraphics[width=1.\linewidth]{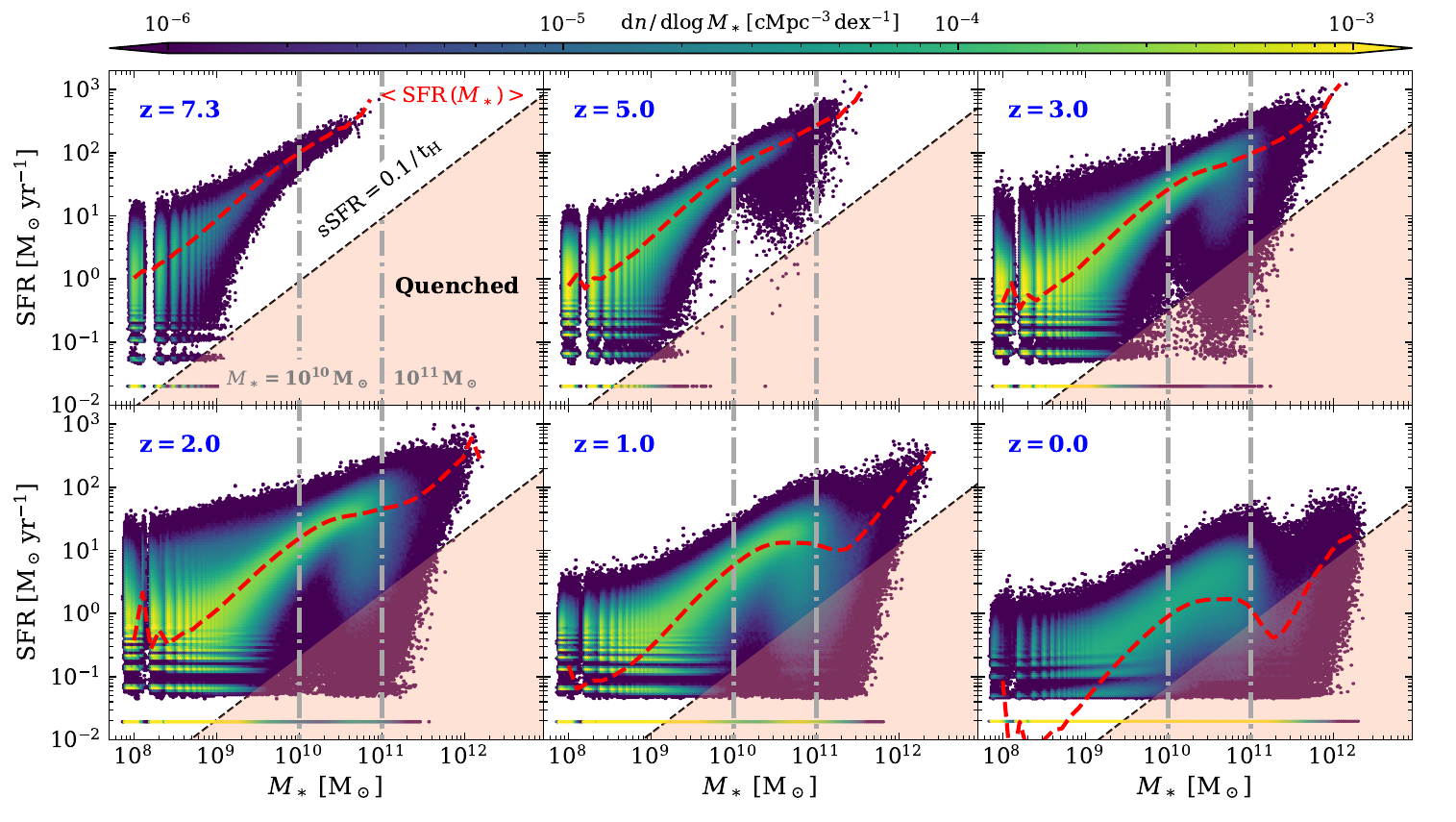}
\vspace{-0.3cm}
\caption{SFR distribution of Sample 0 (which includes all galaxies with non-zero stellar mass) predicted by the FLAMINGO simulation as a function of stellar mass at various redshifts. The black dashed lines indicate sSFR$\,{=}\,0.1/t_{\rm H}(z)$ (see Eq.~\ref{eq_quenched}), below (above) which we define as quenched (star-forming) galaxies. The galaxies with SFR below the numerical resolution, thus assigned zero values, were given arbitrary values of $2\times10^{-2}\,{\rm M_\odot\,yr^{-1}}$ for the presentation. The red dashed lines show the average SFR as a function of stellar mass, including those with zero SFR. The mass range of our main sample used for most of the later analysis, `Main Sample', defined as those with stellar mass between $10^{10}\,{\rm M}_\odot$ and $10^{11}\,{\rm M}_\odot$, is indicated by the gray dot-dashed lines. The stripe patterns in both the low-mass and low-SFR end are due to the limit of numerical resolution where a galaxy consists of a few baryonic particles.   }
\label{fig_SFMS}
\end{figure*}

First, we define `Sample 0', the most inclusive sample we construct, which contains all simulated galaxies at any redshift with non-zero stellar mass assigned by FLAMINGO. This is an inevitable selection because the aim of our study focuses on the quenching of galaxies, which is only defined at given stellar mass. Specifically, we define quenched galaxies as those satisfying the following condition: 
\begin{align} \label{eq_quenched}
{\rm sSFR}\cdot t_{\rm H}(z) < 0.1, 
\end{align}
where ${\rm sSFR}$ is the specific SFR, namely the SFR divided by the stellar mass, and $t_{\rm H}(z)$ is the Hubble time. ${\rm sSFR}\cdot t_{\rm H}(z)$ is the number of mass $e$-folds during the Hubble time when assuming a constant sSFR: 
\begin{gather}
\alpha\equiv{\rm sSFR}\cdot t_{\rm H}(z) = \dd M_\ast/\dd t/ M_\ast \cdot t_{\rm H}(z), \nonumber \\
M_\ast(t+t_{\rm H}(z)) = M_\ast(t) \cdot e^{\alpha}.
\end{gather}
This criterion of quenched galaxies is broadly consistent, albeit slightly tighter, with what has been considered in the literature, also generally thought to match roughly the traditional color-based selection \citep[e.g.,][]{Gallazzi2014, Pacifici2016, Leja2019, RM2019, Carnall2024, Baker2025}. However, even lower values for the criteria have also been adopted in some studies \citep[e.g.,][]{Park2022, Tacchella2022}, making our choice roughly an average. As can be seen from the equation, this criterion selects quiescent galaxies that do not increase their stellar mass by 10 per cent or more over the next Hubble time, assuming a constant SFR. We do not find significant changes in our results and conclusions when changing the criterion by a factor of a few, or using the SFR averaged over a period, e.g., of 100\,Myr. Sample 0 is used in Fig.~\ref{fig_SFMS}, for instance, where the SFR--$M_\ast$ plane of the galaxies is investigated, as the purpose of the sample is to explore the overall properties and trends predicted by the simulation, as well as to identify outliers, simulation limits, artefacts, or any other unexpected/undefined behaviors. A summary of the properties of Sample 0 is provided in Table~\ref{tab_sample}. 

\begin{table}
 \renewcommand{\arraystretch}{1.2} 
 \centering
  \begin{minipage}{90mm}
  \caption{Summary of the sample properties.}
  \begin{tabular}{llr}
\hline
Sample & Selection & \# of galaxies\textsuperscript{\footnote{At $z\,{=}\,0$.}}\\
\hline
\hline
Sample 0 & $M_\ast\,{>}\,0$ & 44,514,486 \\
Main Sample & $M_\ast\,{=}\,[10^{10},10^{11}]\,{\rm M}_\odot$ & 5,297,737 \\
Main Sample-lowBH & $M_{\ast}\,{=}\,[10^{10},10^{11}]\,{\rm M}_\odot$ & 3,624,846 \\
 & $M_{\rm BH}\,{\leq}\,10^7\,{\rm M}_\odot$ &  \\
\hline
\\
\vspace{-8mm}
\end{tabular}
\textbf{Notes.}
\vspace{-5mm}
\label{tab_sample}
\end{minipage}
\vspace{-0.2cm}
\end{table}

Sample 0 also helps to set the range of mass reliable for our analysis, which leads to construction of another sample, the `Main Sample', which is the main sample for our analysis, also the second most inclusive sample following Sample 0. Specifically, from Fig.~\ref{fig_SFMS}, we identify the stellar mass of $\,{\simeq}10^{10}\,{\rm M}_\odot$ where the lower limit of SFR due to the numerical resolution falls below the quenching criterion of Eq.~\ref{eq_quenched}, i.e. the black dashed line. Below this stellar mass, the quenched fraction of galaxies at a given mass is not robust, because those with zero SFR values assigned by the simulation falling below the numerical limit may be above or below the criterion, thus quenched or not. Indeed, the quenched fraction, as well as the SFR of the star-forming galaxies, has been shown to be artificially boosted below the mass due to this effect, as discussed in \cite{Schaye2023}. 

While we determine the lower limit of $M_\ast\,{=}\,10^{10}\,{\rm M}_\odot$, detailed investigation of Fig.~\ref{fig_SFMS} also hints that there should be an upper limit in stellar mass for analysis that can be trusted. This is due to the `upturn' in the SFR shown at $M_\ast\,{\gtrsim}\,10^{11}\,{\rm M}_\odot$ at $z\,{\lesssim}\,2$. We will discuss this further in detail in Sect.~\ref{ssec_SFRMs}. 

We therefore construct Main Sample by only including the subset of Sample 0 with the stellar mass between $M_\ast\,{=}\,10^{10}\,{\rm M}_\odot$ and $10^{11}\,{\rm M}_\odot$. As a result, this sample essentially focuses on Milky Way-mass systems. In total, the Main Sample consists of approximately 5.3 million galaxies. Later, as we find that the main drivers for quenching are largely different for galaxies with $M_{\rm BH}\,{\geq}\,10^7\,{\rm M}_\odot$ and $M_{\rm BH}\,{\leq}\,10^7\,{\rm M}_\odot$, we define and probe only a subset of Main Sample with $M_{\rm BH}\,{\leq}\,10^7\,{\rm M}_\odot$, denoted as Main Sample-lowBH and comprising ${\simeq}$\,3.7 million galaxies (or about two thirds of Main Sample), for some of the analysis.

\subsection{Random Forest analysis}
\label{ssec_RF}

Later, to determine the parameters directly correlated with the quiescence of galaxies in different samples, we employ a Random Forest (RF) classifier, a type of machine learning approach widely used to answer a range of classification problems. In our application, the task is a single-label classification with two classes: star-forming and quenched galaxies. RF has proved to be particularly powerful for distinguishing direct and indirect variables, or first-hand and secondary correlation, in its application in astronomy \citep[e.g.,][]{Bluck2020a, Bluck2020b, Bluck2022, Bluck2024, Goubert2024}. RF provides the relative importance of the features, computed as the fractional contribution of each feature to minimizing the Gini coefficient throughout each tree. The Gini coefficient is a measure of impurity in given data, thus minimizing the coefficient means attempting to achieve the best separation/classification at each `node', or `decision fork'. 

In our analysis, following \cite{Bluck2022} (B22, hereafter), we employ the RandomForestClassifier of the scikit-learn \citep{scikit-learn} python package. We set 250 decision trees for each forest, and each tree consists of a maximum of 500 nodes. To avoid an overfitting problem, the minimum number of samples required to be at a leaf node is fixed at 120. We then construct a total of 10 forests, for each of which bootstrapped samples are utilised as the input training and test set, to compute the average and the variance of the results. We checked that varying these hyper-parameters by a factor of a few does not have a significant impact on the results. 

We split the sample for a forest into a training and test set, each containing half. Also, we construct the data for each forest such that the sample contains an equal number of quenched and star-forming galaxies, because otherwise the results can be biased towards over-weighting the parameters responsible for one population over the other \citep{scikit-learn}. In practice, the construction of the balanced sample is achieved by undersampling the more dominant class within the whole sample in each case. We also tried other machine learning techniques to take into account the imbalanced sample, such as SMOTE \citep{Chawla2002} which oversamples the minority class by interpolating between randomly chosen neighbors. The improvements, however, are found insignificant compared to the simple undersampling for our case. This is in line with some recent studies that found the popular algorithms for balancing the data do not improve prediction performance for strong state-of-the-art classifiers in general \citep{ElorAverbuchElor2022}. Thus, we simply choose to undersample the majority class to construct the input data for each of the RF classifiers. 

We set up the RandomForestClassifier to consider all features at every node. B22, in their appendix, have shown that the power of RF for separating direct and indirect variables is significantly reduced when only a subset of features are used at decision forks.

Finally, before running the classifiers, we always rescale all the features to logarithmic values, as well as to have the same medians and quantile ranges, in order to ensure no bias introduced by different dynamic ranges of the features. The set of features chosen for our analysis will be introduced later in Sect.~\ref{ssec_drivers}, motivated by the findings therein. 

\begin{figure*}
\includegraphics[width=1.\linewidth]{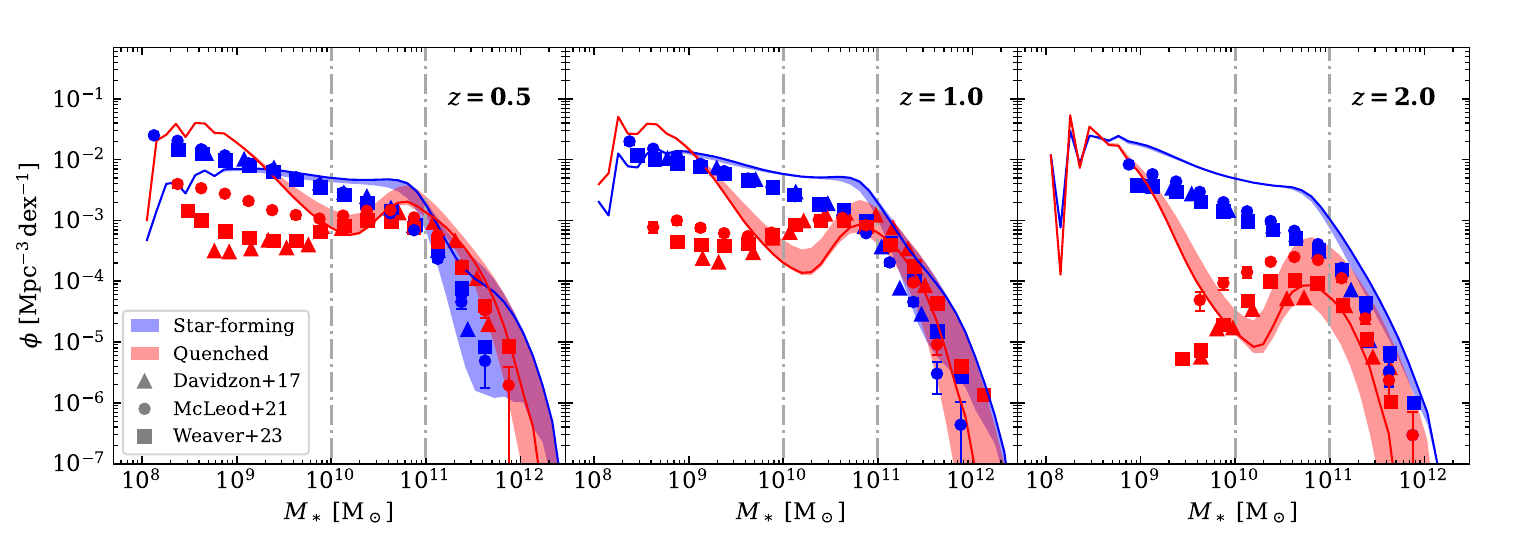}
\vspace{-0.5cm}
\caption{Stellar mass function (SMF) of main sequence (blue) and quiescent (red) galaxies up to redshift of 2 from FLAMINGO. The solid lines show the results using our fiducial cut of sSFR$\,\cdot\,t_{\rm H}(z)\,{=}\,0.1$ to separate main sequence from quenched galaxies, while the bands cover the range of sSFR$\,\cdot\,t_{\rm H}(z)\,{=}\,0.01-1$. For comparison, the observational results of \citet{Davidzon2017}, \citet{McLeod2021}, and \citet{Weaver2023} are included, all of which obtained a set of near-IR-selected SMFs separately for passive and star-forming galaxies, based on a color selection. The mass range of our main sample used for most of the later analysis, `Main Sample', defined as those with stellar mass between $10^{10}\,{\rm M}_\odot$ and $10^{11}\,{\rm M}_\odot$, is indicated by the gray dot-dashed lines. }
\label{fig_SMF}
\end{figure*}

\section[result]{Results}
\label{sec_result}

\begin{figure*}
\includegraphics[width=1.\linewidth]{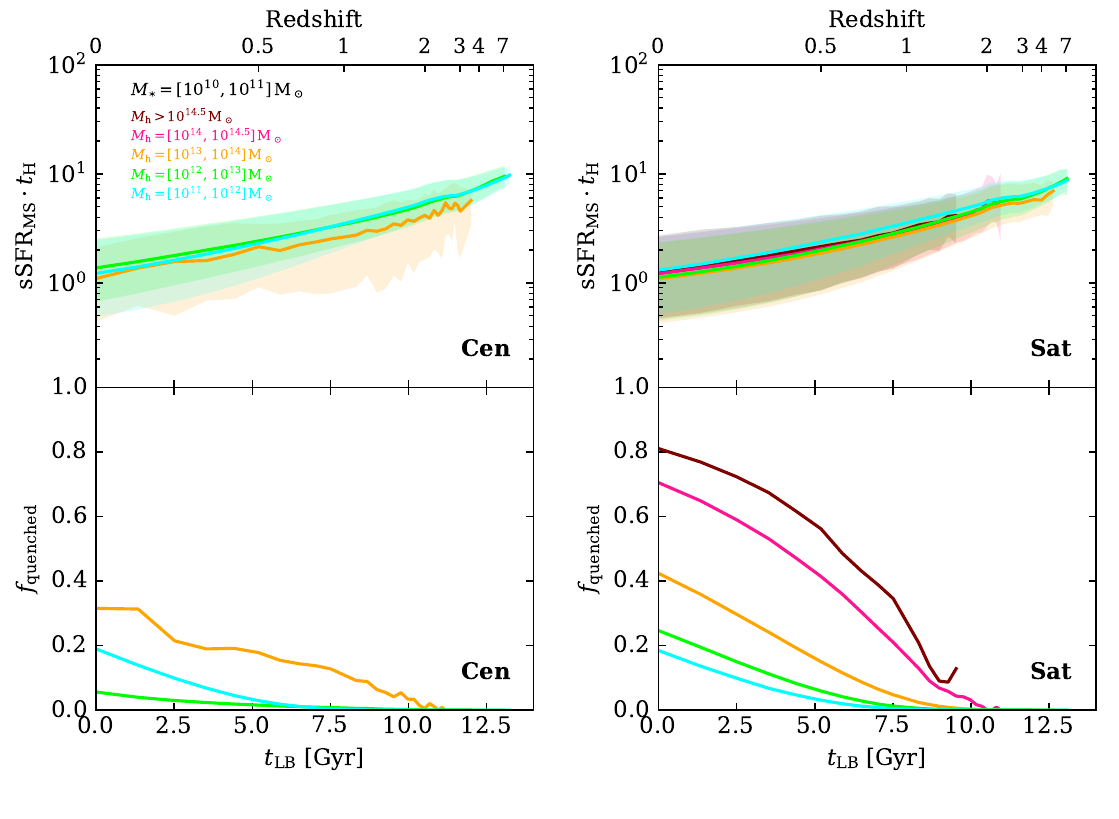}
\vspace{-0.8cm}
\caption{${\rm sSFR}\cdot t_{\rm H}(z)$, which is the number of $e$-folds in stellar mass during the Hubble time assuming a constant sSFR at its current rate, of the galaxies on the star-forming main sequence (defined as those with ${\rm sSFR}\cdot t_{\rm H}(z)>0.1$; upper panels), and the fraction of quenched galaxies (those with ${\rm sSFR}\cdot t_{\rm H}(z)<0.1$; lower panels), as predicted by the FLAMINGO simulation. The medians and 16th-84th percentiles in haloes of different mass are indicated by the curves and bands of colors, respectively. Centrals (left panels) and satellites (right panels) of stellar mass between $10^{10}$ and $10^{11}\,{\rm M}_\odot$ are selected independently from each of the snapshots, thus not tracking the same objects. Note that the main-sequence galaxies have remarkably similar sSFR in haloes of different mass and even between centrals and satellites, while the quenched fraction is sensitive to environment. }
\label{fig_sSFHfqz}
\end{figure*}

\subsection{The SFR--$M_\ast$ plane}
\label{ssec_SFRMs}

As explained in Sect.~\ref{sec_methods}, we first investigate all galaxies with non-zero stellar mass assigned by FLAMINGO, namely Sample 0. Fig.~\ref{fig_SFMS} shows the density map of its distribution on the SFR--$M_\ast$ plane from $z\,{\simeq}\,7$ to 0, together with the mean SFR as a function of stellar mass, denoted by the red dashed line. As already discussed in Sect.~\ref{sec_methods}, note that below $M_\ast\,{\simeq}\,10^{10}\,{\rm M}_\odot$ there are galaxies whose SFR fall under the limit dictated by the numerical resolution of FLAMINGO. Since this leads to bias in the identification of quenched and star-forming galaxies, we decide to eliminate the galaxies of $M_\ast\,{\leq}\,10^{10}\,{\rm M}_\odot$ from our analysis. The red line showing the mean SFR$(M_\ast)$, however, is calculated including those with zero SFR values. 

For those systems of $M_\ast\,{\geq}\,10^{11}\,{\rm M}_\odot$, we identify an upturn in the SFR (as traced by the line of the mean SFR) whose origin is not certain. Whether this upturn is due to the miscentering of BHs even after the attempt of repositioning, or the subgrid model for AGN feedback is unclear. Most observations probe the SFR--$M_\ast$ relation only up to $M_\ast\,{\simeq}\,10^{11.2}\,{\rm M}_\odot$ directly, while extensions to higher mass through empirical fittings normally do not show such upturn \citep[e.g.,][]{Speagle2014, Tomczak2016, Leja2022}. Also, FLAMINGO only has their free model parameters calibrated to the SMF at $M_\ast\,{\lesssim}\,10^{11.5}\,{\rm M}_\odot$, ignoring the higher mass regime due to systematics between different data set, choices of aperture, and treatments of intracluster light. Despite that, however, the simulation predictions are found to match the BH mass from observations at $M_\ast\,{\gtrsim}\,10^{11.5}\,{\rm M}_\odot$, as shown in \cite{Schaye2023}. 

The development of the low-SFR tail for the $M_\ast\,{\gtrsim}\,10^{10}\,{\rm M}_\odot$ systems seen at $z\,{\simeq}\,5$ and later is believed to be due to the AGN feedback. Thus the Main Sample also explores the transitioning regime, well suited for studying the impact of AGN on quenching and the galaxy evolution. 

\subsection{Stellar mass function of main sequence and quenched galaxies}
\label{ssec_SMF}

Following the inspection of the SFR--$M_\ast$ plane, we examine the galaxy stellar mass function (SMF) of star-forming and passive galaxies separately, as defined by the criterion in Eq.~\ref{eq_quenched}. The SMFs are shown in Fig.~\ref{fig_SMF}, compared with the observational results of \citet{Davidzon2017}, \citet{McLeod2021}, and \citet{Weaver2023}. \citet{Davidzon2017} and \citet{Weaver2023} estimated near-IR (nIR) selected SMFs up to redshift of 6--7, in the COSMOS field. Similarly, \citet{McLeod2021} also derived nIR SMFs over $0.25\,{\leq}\,z\,{\leq}\,3.75$ by combining ground-based and Hubble Space Telescope data. However, all these studies, and most other similar works, classified passive and star-forming galaxies relying on a color cut across all redshifts, whereas ours is based on sSFR$\,\cdot\,t_{\rm H}(z)$, which also evolves with redshift. Furthermore, various combinations of color planes and cuts have been adopted among the studies. This difference complicates a direct comparison. To address this, we present SMFs using a range of sSFR$\,\cdot\,t_{\rm H}(z)\,{=}\,0.01-1$ to define main-sequence and quiescent galaxies, alongside those obtained with our fiducial threshold of sSFR$\,\cdot\,t_{\rm H}(z)\,{=}\,0.1$. The definition of passive and star-forming galaxies proves critical, particularly for the quiescent population at high redshifts, altering their SMF by factors of 2 (at $z\,{=}\,0.5$) to 10 (at $z\,{=}\,2$). The agreement with observations varies significantly depending on the chosen threshold. Despite these uncertainties, the FLAMINGO simulation reproduces the observed trend of an decreasing quiescent fraction with redshift. While the average quenched fraction within the mass range of our Main Sample is underestimated, the mass scale at which the quenched and star-forming SMFs cross, which is more relevant for our later analysis and interpretations, always remains within about a factor of 2 of the observations. Additionally, the impact of the selection threshold grows substantially at higher redshifts, which may partly explain the larger discrepancies there. As noted later in Sect.~\ref{ssec_drivers}, our key results are insensitive to the choice of the threshold, due to the rapid transition from main sequence to quiescence, while such changes can improve the agreement between the observations and the model prediction of SMF.

\subsection{Main sequence and quenched galaxies in different environments}
\label{ssec_MSQ}

Now we investigate dependence of the quiescent and main-sequence galaxies on environment. As a proxy of the galaxy environment, here we adopt halo mass, $M_{\rm h}\,{=}\,M_{200}$. We reach similar conclusions, however, when making use of other environment indicators such as $M_{\rm \ast,1.5pMpc}$, the total stellar mass within a 3-D sphere of 1.5\,pMpc, to probe the dependence. 

Fig.~\ref{fig_sSFHfqz} presents the median and 16--84 percentile range of ${\rm sSFR}\cdot t_{\rm H}(z)$ for the star-forming galaxies in the upper panel, while the fraction of quenched galaxies is shown in the lower panel. The results are shown for five halo mass bins as indicated by the different colors. The selections of systems are made independently at each snapshot, i.e. the curves are not tracking the same objects across redshifts. 

A striking fact from Fig.~\ref{fig_sSFHfqz} is that, for satellite galaxies, the star-forming ones show no environmental dependence of their sSFR while their quenched fraction is a strong function of the host halo mass, with those in more massive haloes more likely to be quenched. A similar trend is found even for centrals. Even more strikingly, the centrals and satellites on the star-forming main sequence are almost indistinguishable. Combining all these results, non-quiescent galaxies practically share the same star-formation efficiency regardless of their environments as well as of their local dominance, i.e., whether they are centrals or not. This is consistent with the observational finding of \citet{Peng2012}. However, the quenched fraction varies significantly according to the environments and local dominance at any given time in cosmic history. As will be discussed in Sect.~\ref{sec_discussion}, these results indicate a rapid, abrupt transition from star-forming to quiescent state, implying a short timescale of less than a Gyr and a remarkable effectiveness of the physical mechanism responsible for quenching, as well as its dependence on environment.

\begin{figure*}
\includegraphics[width=0.93\linewidth]{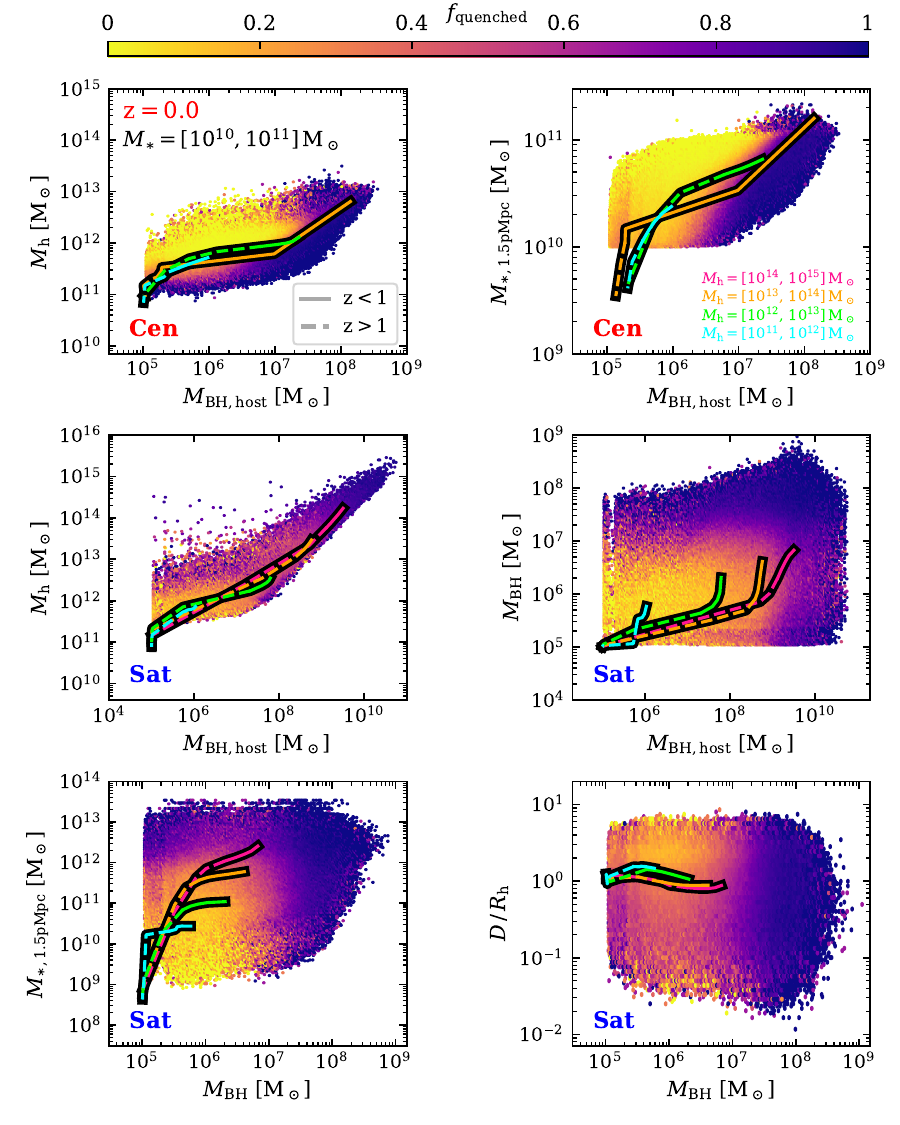}
\caption{The quenched fraction of galaxies projected on multiple 2D parameter spaces of galaxy and environment properties, from the $z\,{=}\,0$ snapshot of FLAMINGO (`Main Sample'). The properties include host halo mass ($M_{\rm h}$), black hole (BH) mass ($M_{\rm BH}$), BH mass of the host halo ($M_{\rm BH, host}$), stellar mass integrated within a 3D sphere of 1.5\,pMpc centered on the galaxy in question ($M_{\rm \ast, 1.5pMpc}$), and distance from the center of host halo divided by halo radius ($D/R_{\rm h}$). Centrals are shown in the top panels, while the rest show the results for satellites. The thick curves represent the median histories of galaxies with $M_{\ast, z{=}0}$ between $10^{10}$ and $10^{11}\,{\rm M}_\odot$ at $z\,{>}\,1$ (dashed) and $z\,{<}\,1$ (solid). Clearly, galaxy quenching as predicted by FLAMINGO is a cooperation between intrinsic (or in-situ) and external (ex-situ) mechanisms, with the former dominating at $M_{\rm BH}\,{\gtrsim}\,10^7\,{\rm M}_\odot$ while the environmental effects play a role at $M_{\rm BH}\,{\lesssim}\,10^7\,{\rm M}_\odot$. }
\label{fig_fq_ND}
\end{figure*}

\begin{figure*}
\includegraphics[width=1.01\linewidth]{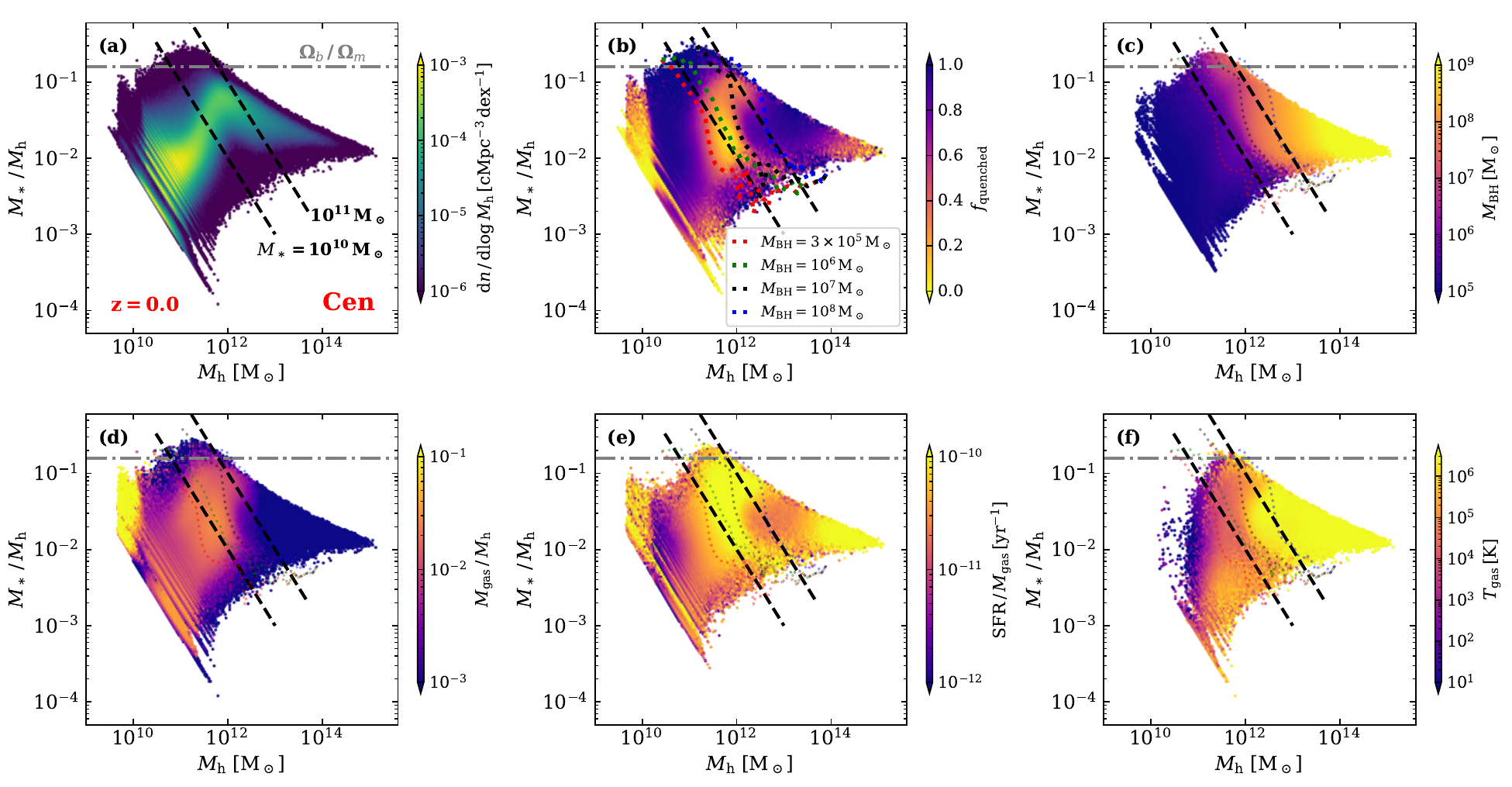}
\caption{Central galaxies on the $z\,{=}\,0$ stellar mass--halo mass (SMHM) plane colour-coded by different properties in the different panels. In all panels, the black dashed lines indicate our main sample with stellar masses between $10^{10}$ and $10^{11}\,{\rm M}_\odot$. The gray dot-dashed line represents the cosmic baryon fraction. The galaxies above the cosmic fraction are mostly systems that were once satellites in the past during which they lost dark matter mass. (a) The colour-coding corresponds to the number density distribution of central galaxies. (b): Quenched fraction. It is seen that galaxies are least quenched around the Milky Way-mass halo of $M_{\rm h}\,{\simeq}\,10^{12}\,{\rm M}_\odot$, while the quenched fraction rapidly increases in haloes of $M_{\rm h}\,{\gtrsim}\,10^{12.5}\,{\rm M}_\odot$. The dotted line indicates several BH mass (repeated in the subsequent panels), with the black line representing $M_{\rm BH}\,{=}\,10^7\,{\rm M}_\odot$, above (below) which the galaxies are mostly quenched (active) at $M_{\rm h}\,{\gtrsim}\,10^{12}\,{\rm M}_\odot$. This suggests that BH feedback is the responsible mechanism for quenching centrals in these haloes. (c): BH mass on the SMHM plane. The dotted lines represent the same lines of the selected BH masses from (b), to guide the eye. (d): Gas mass (in any phase) within 50\,pkpc divided by halo mass. (e): Star-formation efficiency (SFE), defined as SFR within 50\,pkpc divided by the gas mass. (f): Gas temperature averaged over all gas in any phase within 50\,pkpc. The bottom three panels demonstrate that the sudden quenching seen in (b) near $M_{\rm BH}\,{=}\,10^7\,{\rm M}_\odot$ is driven by a combination of plummeted SFE due to the AGN heating of gas, and deficit of gas in the reservoir, both of which trace the CGM amount. }
\label{fig_fq_Mh}
\end{figure*}

\subsection{Ex-situ vs. in-situ drivers of quenching}
\label{ssec_drivers}

Here we investigate a more complete set of both ex-situ and in-situ indicators that may be responsible for the quenching of galaxies. Also by taking advantage of the RF, we identify the physical properties that are more directly correlated with the quiescence, separately from secondary variables that only present correlations via their dependence on the primary parameters. 

\subsubsection{Quenched fraction as a function of parameters}

Fig.~\ref{fig_fq_ND} projects the quenched fraction of the Main Sample at $z\,{=}\,0$ on the 2-D planes of various properties, including the host halo mass ($M_{\rm h}\,{=}\,M_{200}$), the BH mass ($M_{\rm BH}$), the host BH mass ($M_{\rm BH, host}$), the total stellar mass within 1.5\,pMpc ($M_{\rm \ast,1.5pMpc}$), and the distance from the host halo center (normalized to $R_{\rm h}\,{=}\,R_{200}$, i.e. $D/R_{\rm h}$). For centrals, the host BH mass is the same as their BH mass, which is the subgrid mass of the most massive BH found within 50\,pkpc from the center of a galaxy. For satellites, the host BH mass is $M_{\rm BH}$ of their central galaxy. The mean evolution track for a given present-day halo mass is given by the lines of different colors. For these lines, the same objects of given present-day mass have been tracked using the merger tree. 

Most notably, there are two regimes with different characteristics, one is $M_{\rm BH}\,{\gtrsim}\,10^7\,{\rm M}_\odot$ where basically every galaxy is quenched, while at $M_{\rm BH}\,{\lesssim}\,10^7\,{\rm M}_\odot$ the impact of secondary parameters of environment indicators such as the host halo mass, $M_{\rm \ast,1.5pMpc}$, and the properties related to the host, are evident. This is true for both centrals and satellites, while the dividing line of $M_{\rm BH}\,{\simeq}\,10^7\,{\rm M}_\odot$ is clearer for satellites. For centrals, the transition to quiescence is a combined function of the BH mass and the environment, with a higher threshold BH mass for quenching in more massive haloes, which is as expected due to higher cooling rate in those haloes \citep[see also e.g.,][]{Bluck2022, Goubert2024}. A remarkable thing to note is how sharp and abrupt the transition between star-forming and quenched galaxies is, once crossing the dividing line in $M_{\rm BH}$. This is consistent with the finding of Fig.~\ref{fig_sSFHfqz} and the previous subsection, both indicating that the impact of AGN feedback on quenching is rapid and sudden, which indeed is confirmed to be the case as will be discussed in Sect.~\ref{sec_discussion}. Note that the characteristic BH mass where the transition occurs is insensitive to the choice of the threshold set to 0.1 in Eq.~\ref{eq_quenched} to define the quenched and star-forming galaxies, because the transition is so sharp and rapid, taking place over a very narrow range in BH mass. Therefore finding the critical BH mass at $M_{\rm BH}\,{\simeq}\,10^7\,{\rm M}_\odot$ is not due to any of the assumptions applied by our analysis, but a genuine prediction of the model. Interestingly, this critical mass is consistent with BH mass at which galaxies are found predominantly quenched in observations \citep[e.g.,][]{Terrazas2016, Bluck2020b, Chen2020, Piotrowska2022}. We confirm that the transition BH mass barely changes when the threshold is altered by a factor of 5 or even more in either direction. 

The trend and strength of the dependence on the various environmental proxies also vary. For central galaxies, the halo mass is shown to have a stronger quenching effect at a given BH mass, than $M_{\rm \ast,1.5pMpc}$. Fig.~\ref{fig_fq_Mh} shows this more clearly. The top-left panel presents the number density distribution of both the whole and our sample galaxies (indicated by the dashed lines) selected from FLAMINGO on stellar mass--halo mass plane. Only central galaxies are chosen for the presentation. The top-middle panel of Fig.~\ref{fig_fq_Mh}, on the other hand, shows the quenched fraction on the same plane. Going from left to right across the figure, central galaxies are clearly seen to be least quenched within the Milky Way-mass haloes of $M_{\rm h}\,{\simeq}\,10^{12}\,{\rm M}_\odot$, while the quenched fraction increases rapidly in haloes of both $M_{\rm h}\,{\lesssim}\,10^{11}\,{\rm M}_\odot$ and $M_{\rm h}\,{\gtrsim}\,10^{12.5}\,{\rm M}_\odot$\footnote{Note that those above the cosmic baryon fraction (the gray dashed line in both panels) are, in most cases, systems that were once satellites when significant fractions of their dark matter mass were lost via dynamical processes.}. The quenched fraction at the low-mass end of $M_\ast\,{\lesssim}\,10^{9.5}\,{\rm M}_\odot$, however, is likely over-estimated due to the limit of numerical resolution, as it is the mass regime that the lowest possible SFR value from FLAMINGO stays above our definition of quiescence (Fig.~\ref{fig_SFMS}; see also \citealt{Schaye2023}), which is the reason we excluded them to define Main Sample. Notably, the dotted line of $M_{\rm BH}\,{=}\,10^7\,{\rm M}_\odot$ aligns sharply with the divide between the quiescent and main-sequence population at $M_{\rm h}\,{\gtrsim}\,10^{12}\,{\rm M}_\odot$, strongly suggesting that BH feedback is the main quenching mechanism for centrals in those haloes. This is because the BHs in FLAMINGO grow rapidly (over less than 2\,Gyr span) from its seed mass of $M_{\rm BH}\,{\simeq}\,10^{5}\,{\rm M}_\odot$ up until $\,{\simeq}\,10^{7}\,{\rm M}_\odot$, and so does the impact of AGN feedback. 

Yet another question is whether the AGN feedback ceases the star formation by ejecting the galactic gas to the CGM \citep{Maiolino2012, Harrison2014} or further out \citep{Davies2019, Davies2020, Oppenheimer2020}, or by heating the gas too hot to form stars \citep{Fabian2012, PengMaiolinoCochrane2015, Zinger2020}, namely if it is lack or the state of the gas reservoir that quenches galaxies \citep{Genzel2015, LinL2017, LinL2020, Tacconi2018, Ellison2020, Ellison2021, Dou2021}. The bottom panels of Fig.~\ref{fig_fq_Mh} explore this, presenting the gas mass divided by halo mass (including all gas in any phase within 50\,pkpc), star-formation efficiency (SFE; defined as the SFR divided by the gas mass) within 50\,pkpc, and the mean temperature of gas (in any phase within 50\,pkpc). Being larger than a typical size of galaxies while still smaller than the virial radius, 50\,pkpc aperture probes the inner CGM. Interestingly, all three panels reveal sharp transitions near $M_{\rm BH}\,{\simeq}\,10^{7}\,{\rm M}_\odot$ (the black dotted line), similar to what is seen for the quiescent fraction: the gas mass fraction decreases roughly by a factor of 3, the SFE by a factor of 10, while $T_{\rm gas}$ increases by a factor of 100, across $M_{\rm BH}\,{\simeq}\,10^{7}\,{\rm M}_\odot$. This suggests that the sudden quenching of galaxies around the threshold BH mass is primarily driven by plummeted SFE, which in turn is due to the gas reservoir heated too hot to further star formation, while deprivation of the reservoir also has a significant impact. However, it is important to note that, by design, the simulation enforces a strong dependence of SFR on the ISM gas mass. As a result, what the SFE effectively traces is the ratio of the inner CGM to the ISM mass, which can be influenced by consumption of the ISM due to star formation, by ejection of the ISM into the inner CGM, by ejection of the inner CGM into the outer region, and by preventive impact on the CGM not to accrete on to the ISM (see also \citealt{MitchellSchaye2022}). We find that the same three quantities of the outer CGM (all bound gas outside 50\,pkpc) present much smoother transitions across the threshold BH mass, which thus cannot explain the sudden increase in the quiescent fraction. This prediction for the combined quenching effect of reduced gas fraction and SFE is consistent with findings from some recent observational studies \citep[e.g.,][]{Tacconi2018, Colombo2020, Ellison2020, Piotrowska2020, Dou2021b}. The strong dependence on $T_{\rm gas}$, on the other hand, indicates the transition from the cold to the hot accretion regime. We confirm the same conclusion with RF analysis where over 95 per cent relative importance for predicting the galaxy quenching is given to the SFE and gas fraction of the inner CGM, both of which trace the CGM amount, as will be discussed in Sect.~\ref{ssec_RFa}. This is consistent with some other state-of-the-art simulations such as EAGLE, IllustrisTNG, and Illustris \citep{Piotrowska2022}. 

In Fig.~\ref{fig_fq_ND}, the trends of quiescent fraction with environments are much less clearly shown for $M_{\rm \ast,1.5pMpc}$ than for halo mass. At the low-mass end, this can be understood that, due to the low baryonic mass resolution of FLAMINGO, the environment of the low mass systems are not properly traced by $M_{\rm \ast,1.5pMpc}$. This is evidenced by the sharp lower-limit cut at $10^{10}\,{\rm M}_\odot$ on the $M_{\rm \ast,1.5pMpc}$ axis in the sample distribution in the top-right panel, which agrees with the stellar mass cut of Main Sample. This means that these are systems with only centrals. At the massive end, due to the shallow slope in the stellar mass--halo mass relation, the dynamic range for $M_{\rm \ast,1.5pMpc}$ is much more compressed than that for $M_{h}$, making it difficult for a clear trend to be revealed. 

\subsubsection{Importance of drivers from Random Forest analysis}
\label{ssec_RFa}

Fig.~\ref{fig_fq_ND} demonstrates that environment also plays an important role in the satellite quenching. Here, however, we attempt to examine their relative importance more quantitatively, which is necessary to reveal main drivers for quenching satellites. To this end, and also for a better assessment of the correlations for centrals, we employ the RF classifier as described in Sect.~\ref{ssec_RF}. The input features for the classifiers are 
\begin{eqnarray} \label{eq_features}
&&\{\boldsymbol{M_\ast, M_{\rm h}, M_{\rm BH}, M_{\rm \ast, 1.5pMpc}, c}, M_{\rm BH, host}, \nonumber \\ 
&&\hspace{2mm}{\rm sSFR_{host}}, D/R_{\rm h} \}, \nonumber
\end{eqnarray}
where those in bold are for centrals, while all features are used for satellites. Although not included in Fig.~\ref{fig_fq_ND}, note that the feature set has additional parameters such as the halo concentration proxy, $c\,{=}\,R_{200}/R_{2500}$, where $R_{2500}$ is the radius within which the average density is 2500 times the critical density, as well as ${\rm sSFR_{host}}$. ${\rm sSFR_{host}}$ is nothing but sSFR but for the central galaxy in case of a satellite. We include ${\rm sSFR_{host}}$ to see if there is any galaxy conformity, namely co-evolution between centrals and satellites in the same environment. Finally, random numbers, `Rand', are included to assess the noise in case of no correlation. 

\begin{figure*}
\includegraphics[width=1\linewidth]{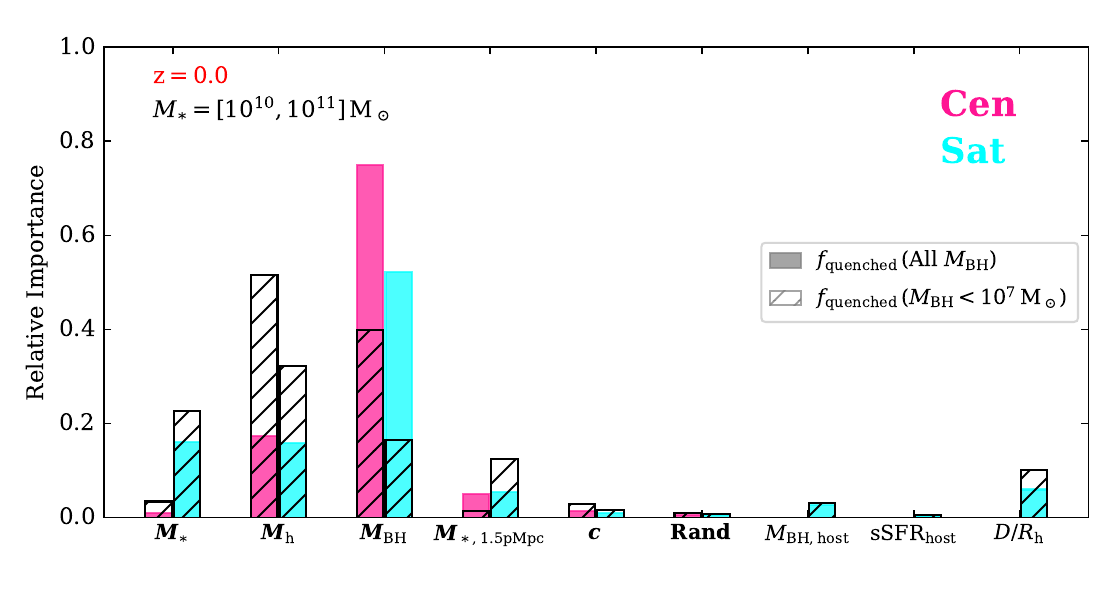}
\vspace{-0.8cm}
\caption{Relative importance of in-situ and ex-situ parameters for quenching, determined from the Random Forest (RF) analysis for the Main Sample (filled bars) and Main Sample-lowBH (hatched bars) from the $z\,{=}\,0$ snapshot of FLAMINGO. The features include stellar mass ($M_\ast$), host halo mass ($M_{\rm h}$), black hole (BH) mass ($M_{\rm BH}$), stellar mass integrated within a 3D sphere of 1.5\,pMpc centered at a galaxy in question ($M_{\rm \ast, 1.5pMpc}$), halo concentration proxy of $R_{200}/R_{2500}$ ($c$), random number (Rand), BH mass of host halo ($M_{\rm BH, host}$), specific star-formation rate of central galaxy (${\rm sSFR_{host}}$), and distance from the center of host halo divided by halo radius ($D/R_{\rm h}$). The features for centrals are in boldface. Clearly, $M_{\rm BH}$ is the most important parameter for galaxy quenching, while halo mass, an environment proxy, also plays an important role at $M_{\rm BH}\,{\lesssim}\,10^7\,{\rm M}_\odot$, where BHs have not yet accumulated enough energy to inject into galaxies to suppress star formation. }
\label{fig_RF_all}
\end{figure*}

Fig.~\ref{fig_RF_all} shows the relative importance between the features for the Main Sample, analyzed by the RF. For both centrals and satellites, the BH mass is the most impactful factor, thus indicating that the in-situ mechanism dominates quenching of galaxies. For centrals, the second dominant parameter is the halo mass, which was already hinted in Fig.~\ref{fig_fq_ND}. The other three parameters have almost negligible impact for centrals. The reason why stellar mass has such little impact is probably because we control the stellar mass of the samples to a narrow dynamic range. $M_{\rm \ast,1.5pMpc}$ is also shown to have a minor effect, part of which is likely due to the reasons discussed above for Fig.~\ref{fig_fq_ND}. Another reason would be that the halo mass is a more direct physical quantity that is related to the dynamics governing the quenching such as the virial shock heating and radiative cooling, over the integrated stellar mass. Inclusion of the halo mass thus reduces the relative importance given to the total stellar mass, as expected. From a test where $M_{\rm h}$ was deliberately removed from the input set, we indeed found that the previous importance given to $M_{\rm h}$ is redistributed to $M_\ast$, $M_{\rm \ast,1.5pMpc}$, and concentration, with almost equal shares among the three. Similarly, including stellar mass also takes away the importance of $M_{\rm \ast,1.5pMpc}$, as for many of the low-mass systems $M_{\rm \ast,1.5pMpc}\,{\simeq}\,M_\ast$. In fact, when $M_\ast$ is removed from the input, it is found that the importance score of $M_{\rm \ast,1.5pMpc}$ increases roughly by that amount. In general, as discussed by e.g. B22, the interpretation of the features given small importance should be cautioned. Another lesson from these results, although obvious, is that a physically responsible driver should be included in the input set of features, in order to be identified so by the classifier. Finally, such small importance given to the random number feature, proves that the RF works as expected, and that our sample is large enough to allow to identify even features of small importance, taking advantage of the large volume of FLAMINGO.

\begin{figure}
\includegraphics[width=1.01\linewidth]{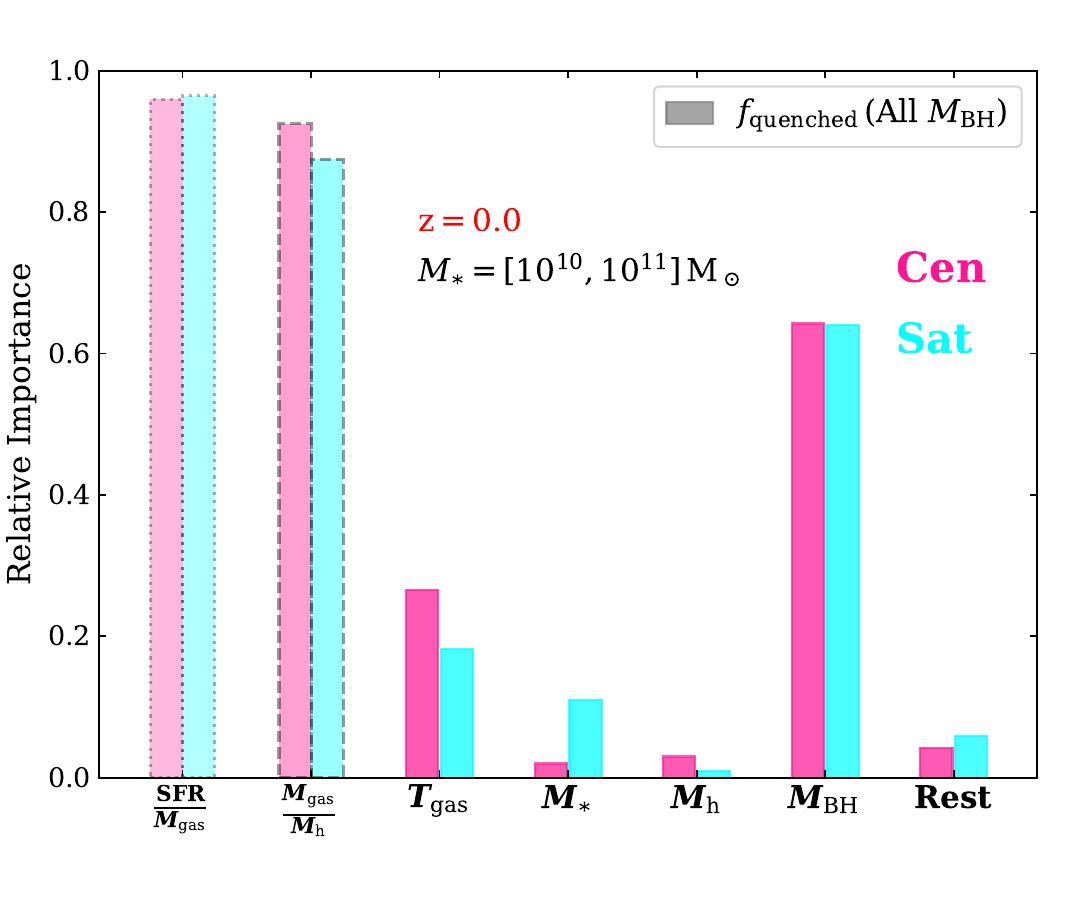}
\vspace{-0.8cm}
\caption{Relative importance of in-situ and ex-situ parameters for quenching, determined from the Random Forest (RF) analysis for the Main Sample from the $z\,{=}\,0$ snapshot of FLAMINGO. Compared with Fig.~\ref{fig_RF_all}, here we further include the SFE (SFR divided by gas mass), gas fraction (gas mass divided by halo mass) and gas temperature, $T_{\rm gas}$, in addition to the feature set. Although the gas properties presented here are those of the inner CGM (gas within 50\,pkpc), we confirm similar results for the outer CGM (all bound gas outside 50\,pkpc). The dotted bars indicate the RF result when all features are considered, while the dashed bars show the result when the SFE is excluded. The other bars result when excluding both SFE and gas fraction. This indicates that the SFE and gas fraction dominate the importance when included, making all the other features nuisance. The features include ${\rm SFR}/M_{\rm gas}$, $M_{\rm gas}/M_{\rm h}$, $T_{\rm gas}$, stellar mass ($M_\ast$), host halo mass ($M_{\rm h}$), black hole (BH) mass ($M_{\rm BH}$), and the sum for the rest of the parameters (namely, stellar mass integrated within a 3D sphere of 1.5\,pMpc centered at a galaxy in question ($M_{\rm \ast, 1.5pMpc}$), halo concentration proxy of $R_{200}/R_{2500}$ ($c$), random number (Rand), BH mass of host halo ($M_{\rm BH, host}$), specific star-formation rate of central galaxy (${\rm sSFR_{host}}$), and distance from the center of host halo divided by halo radius ($D/R_{\rm h}$)). While $T_{\rm gas}$ certainly takes an important role, as a direct indicator of physical state of gas, $M_{\rm BH}$ being dominant over $T_{\rm gas}$ indicates that galaxy quenching is a cumulative process, rather than depending solely on the current state. }
\label{fig_RF_Tgas_reduced}
\end{figure}

\begin{figure*}
\includegraphics[width=1.01\linewidth]{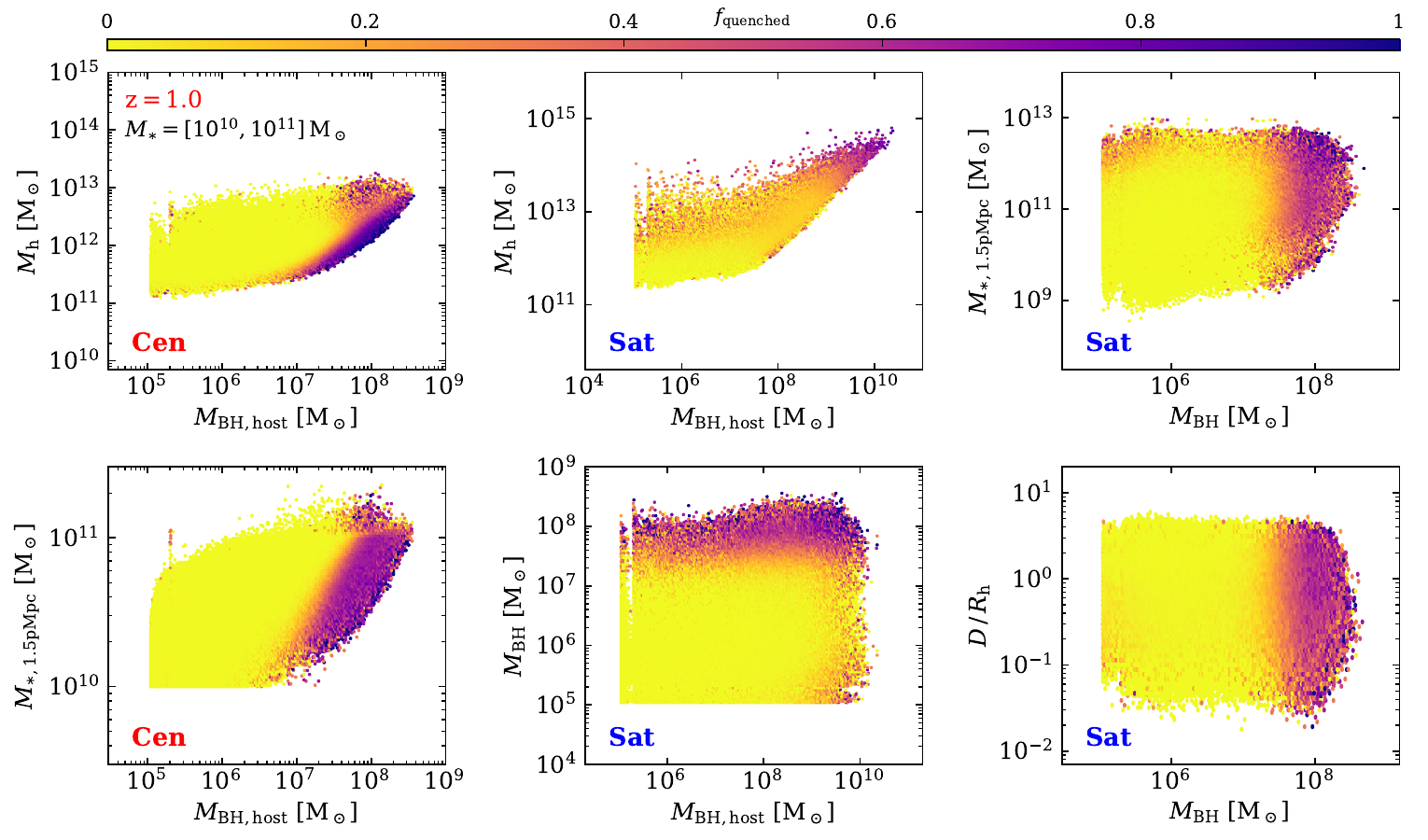}
\caption{The quenched fraction of galaxies projected on multiple 2D parameter spaces of galaxy and environment properties, from the $z\,{=}\,1$ snapshot of the FLAMINGO (Main Sample). The properties include host halo mass ($M_{\rm h}$), black hole (BH) mass ($M_{\rm BH}$), BH mass of the host halo ($M_{\rm BH, host}$), stellar mass integrated within a 3D sphere of 1.5\,pMpc centered at a galaxy in question ($M_{\rm \ast, 1.5pMpc}$), and distance from the center of the host halo divided by halo radius ($D/R_{\rm h}$). Centrals are shown in the left two panels, while the rest show the results for satellites. Compared with Fig.~\ref{fig_fq_ND}, there is still a clear dividing line but at a higher BH mass of approximately $10^{7.5}\,{\rm M}_\odot$ between two distinct regimes, above (below) which quenching is primarily dictated by BH mass (environment).}
\label{fig_fq_ND_z1}
\end{figure*}

\begin{figure}
\includegraphics[width=1.01\linewidth]{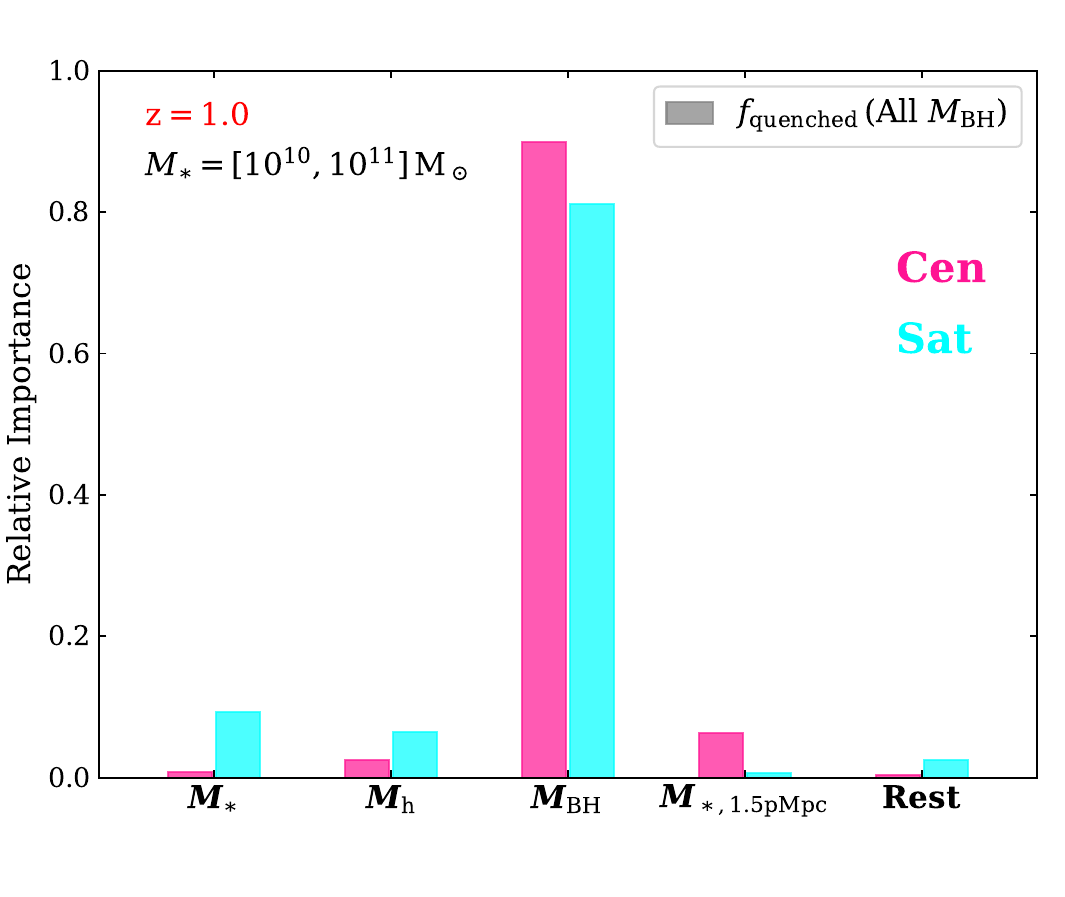}
\vspace{-0.8cm}
\caption{Relative importance of in-situ and ex-situ parameters for quenching, determined from the Random Forest (RF) analysis for all galaxies in Main Sample from the $z\,{=}\,1$ snapshot of FLAMINGO. The features include stellar mass ($M_\ast$), host halo mass ($M_{\rm h}$), black hole (BH) mass ($M_{\rm BH}$), stellar mass integrated within a 3D sphere of 1.5\,pMpc centered at a galaxy in question ($M_{\rm \ast, 1.5pMpc}$), and the sum for the rest of the parameters (namely, halo concentration proxy of $R_{200}/R_{2500}$ ($c$), random number (Rand), BH mass of host halo ($M_{\rm BH, host}$), specific star-formation rate of central galaxy (${\rm sSFR_{host}}$), and distance from the center of host halo divided by halo radius ($D/R_{\rm h}$)). }
\label{fig_RF_all_z1}
\end{figure}

For satellites, the relative significance of BH mass is reduced compared to centrals. Part of this, however, can be attributed to more features being included in the analysis, over which the relative significance is shared. Given that, halo mass has a reduced importance for satellites, while an increased significance is present for stellar mass. This may be understood as halo mass is not a quantity directly related to satellites, while stellar mass is. In case of centrals, the growth of the BH and the impact of the feedback thereof, as well as gravitational potential to maintain cold gas, are related to halo mass. But for satellites, halo mass may be just a larger-scale environment, not reflecting the local gravitational potential as well as the growth of their BHs. There is also a sign of a correlation of the distance to the halo center with quiescence. This may reflect some environment quenching mechanisms such as tidal and ram-pressure stripping that are thought to depend on proximity to the center \citep[e.g.,][]{PintosCastro2019}. No galaxy conformity is demonstrated. This is unsurprising because galaxy conformity should be driven by a fundamental mechanism(s) affecting both centrals and satellites such as assembly history and environments, and itself is unlikely a direct cause of quenching. So, given the nature of RF for discerning direct from secondary correlations, ${\rm sSFR_{host}}$ should be given a negligible importance as shown here, even if there exists a galaxy conformity.

Motivated by the clear dividing line around $M_{\rm BH}\,{=}\,10^7\,{\rm M}_\odot$, across which the physics governing the quenching suddenly seems to change, we do another RF analysis for Main Sample-lowBH, the subset of Main Sample with the lower BH mass of $M_{\rm BH}\,{\leq}\,10^7\,{\rm M}_\odot$. The results are presented in Fig.~\ref{fig_RF_all} as the hatched bars. Compared with the results for Main Sample, the impact of BH is now shown to be much less significant for both centrals and satellites. Most of the previous contribution from BH has now transferred to halo mass. Thus, environment indeed impacts the galaxy quenching for $M_{\rm BH}\,{\lesssim}\,10^7\,{\rm M}_\odot$ where the AGN feedback is expected to be weaker, consistent with the strong environmental dependence from Fig.~\ref{fig_fq_ND}. 

Aside from the parameters in the input set of features, there are some additional properties available in FLAMINGO that are indicative of the physical state of the gas within galaxies and CGM, thus dictating quenching processes more directly. These include the gas fraction (gas mass with respect to halo mass), SFE, and $T_{\rm gas}$ of the inner CGM (gas within 50\,pkpc). We find the RF reveals that the most importance (about 95 per cent for the inner CGM, and slightly higher for the outer CGM) for quenching galaxies is given to SFE when all the three of these gas properties are included, in addition to the other features shown in Fig.~\ref{fig_RF_all}. When excluding the SFE from the analysis, most of the importance (approximately 90 per cent, similarly for the inner and outer CGM) is transferred to the gas fraction, making all the other parameters nuisance (Fig.~\ref{fig_RF_Tgas_reduced}). The SFE being the most important over the gas fraction is consistent with the bottom panels of Fig.~\ref{fig_fq_Mh}, as well as some of other studies using simulations \citep[e.g.,][]{Piotrowska2022}. However, as discussed in Sect.~\ref{ssec_drivers}, both the SFE and gas fraction trace the inner CGM amount, and thus share much of the importance. Interestingly, $T_{\rm gas}$ also becomes important when both the SFE and gas fraction are eliminated from the feature set, taking away quite an importance score that would go to $M_{\rm BH}$ otherwise. This is shown in Fig.~\ref{fig_RF_Tgas_reduced}, which can be compared with Fig.~\ref{fig_RF_all} for the reduced predicting power of $M_{\rm BH}$ for galaxy quenching. Still, the dominance of $M_{\rm BH}$ over $T_{\rm gas}$ may reflect that galaxy quenching is a cumulative process, as is the BH growth, being a result of a long-term energy injection, beyond what the current state of gas such as $T_{\rm gas}$ can trace. Notably, we also find a significant role in quenching attributed to the current BH accretion rate, $\dot{M}_{\rm BH}$, unlike in predictions from some other models \citep{Davies2019, Piotrowska2022}, with a relative importance of up to 0.3 (0.6) for centrals (satellites), comparable to or greater than that for $M_{\rm BH}$. While \citet{Piotrowska2022} also reported a notable importance of 0.1--0.2 given to the BH accretion rate from the EAGLE simulation, the much stronger dependence found here could be due to the relatively poor time resolution of FLAMINGO to capture the instantaneous variability of the BH accretion. 

\subsubsection{Galaxy quenching at higher redshift}

Fig.~\ref{fig_fq_ND_z1} explores the relations on the same projected parameter spaces as in Fig.~\ref{fig_fq_ND} but at $z\,{=}\,1$. Compared to the results at $z\,{=}\,0$, galaxies are overall less quenched. The sharp transition, however, is still identified at a bit higher BH mass of $M_{\rm BH}\,{\simeq}\,10^{7.5}\,{\rm M}_\odot$. This increased threshold BH mass for quenching at higher redshift may be interpreted as a result of the contest between gas cooling and the heating by the AGN: at higher redshifts where the density of gas in galaxy systems is higher so that the cooling is more effective, a higher BH mass must be reached for its AGN feedback energy to be comparable to the faster cooling. Another possible reason could be the higher binding energy at higher redshift \citep[e.g.,][]{BoothSchaye2011}. Below the threshold BH mass of ${\simeq}\,10^{7.5}\,{\rm M}_\odot$, almost all galaxies, both centrals and satellites, are on the star-forming main sequence, with much weakened environmental dependence. This is also demonstrated by the enhanced relative importance of the BH mass up to 0.9, derived from the RF classifier as shown in Fig.~\ref{fig_RF_all_z1}, compared with about 0.7 (0.5 for satellites) at redshift of 0 in Fig.~\ref{fig_RF_all}.

\section[discussion]{Discussion}
\label{sec_discussion}

\subsection{Bimodal evolution and rapid quenching}
\label{ssec_bimodal}

Our results indicate, in Sect.~\ref{ssec_MSQ}, that galaxies of stellar mass between $10^{10}$ and $10^{11}\,{\rm M}_\odot$ follow the same evolution track on the SFR--$M_\ast$ plane, regardless of their environment or whether they are centrals or satellites. Surprisingly, this prediction is found not only at $z\,{\simeq}\,0$ but over the entire cosmic history. At the same time, however, a strong dependence of the quenched fraction on environment is also present across cosmic time. How can these two strong results co-exist and be reconciled? What physical mechanisms or scenarios can explain the significant difference in the evolution of the star-forming main-sequence and quenched galaxies? 

The sudden transition between the two evolutionary trends indicates that the responsible mechanism must have a strong impact on a relatively short timescale, and thus quenching occurs quite rapidly. Otherwise a prolonged tail towards low SFR, which would be a track of galaxies in the process of leaving the main sequence but not quenched yet, should be present. In Fig.~\ref{fig_sSFHfqz}, our results lack any sign of such a tail, implying a sudden quenching. Indeed, we investigated the quenching timescale, $\tau_{\rm quenching}$, defined as the time taken for the number of mass $e$-folds, ${\rm sSFR}\cdot t_{\rm H}(z)$, to fall from 1 to 0.1. Although the distribution of the quenching timescale is found to be broad, a majority of the galaxies are quenched over a relatively short timescale of ${<}\,1.5$\,Gyr. We find that the quenching timescale also depends on the epoch of quenching, with massive galaxies quenched earlier and more slowly. Also, central galaxies are found to have more rapid ($\tau_{\rm quenching}\,{\simeq}\,0.7$\,Gyr) and narrower spread of quenching timescale relative to satellites, with a marginal hint of a bimodal distribution (see also Sect.~\ref{ssec_disc_cen}). As we demonstrated in Sect.~\ref{ssec_drivers}, for $M_{\rm BH}\,{\gtrsim}\,10^7\,{\rm M}_\odot$ where the BH and AGN feedback dominates the quenching of galaxies, basically all galaxies are quenched regardless of their environment. The physical mechanism responsible for such strong environmental dependence as shown in Fig.~\ref{fig_sSFHfqz}, therefore, must be something else than the AGN. According to our RF analysis, this is most related to the mass of the host halo, and also to stellar mass to a degree for satellites. 

There are several known physical mechanisms of environmental quenching that depend moderately to strongly on host halo mass, including tidal stripping \citep{Merritt1984},  ram-pressure stripping \citep{GunnGott1972}, galaxy harassment \citep{Moore1996}, and strangulation \citep{DekelBirnboim2006, PengMaiolinoCochrane2015}. These multiple quenching processes in principle may occur over various characteristic timescales spanning an order or magnitude or more, and conditions to be met for each process to be effective also vary and overlap each other \citep{BoselliGavazzi2006}. This makes it difficult to constrain which process is responsible for a given feature in the star-formation history (SFH) over a certain timescale, and their relative importance. Despite that, in general, the ram-pressure stripping is believed to act over a comparatively rapid timescale of a couple of hundred Myr \citep[e.g.,][]{AbadiMooreBower1999} by sudden and complete gas removal, and quench low-mass galaxies. On the other hand, tidal stripping leads to a slow quenching via a more strangulation-like scenario, also requiring high density ICM which is normally met at the central region of haloes. Other merger/interaction-driven `harassment', which can drive quenching by over-consumption of gas via induced boost of star formation, act over a moderate to rapid timescale, being more effective in quenching massive galaxies at high redshift or in dense regions where galaxy encounters are frequent. Given that our samples probe the relatively high stellar mass range, that the similar trends extend to high redshift, and that the quenching has a strong environmental dependence, the most likely mechanism for the sudden quenching is the galaxy interaction-driven harassment. This is further hinted by the decent amount of importance given by the RF to the distance to the halo center, $D/R_{\rm h}$. We confirm, by a traditional approach of investigating the correlation, that satellites with smaller $D/R_{\rm h}$, namely those nearer to the center, thus in denser environment, are more likely quenched.

\subsection{The main drivers for quenching centrals}
\label{ssec_disc_cen}

While in the previous subsection we mainly discussed the finding of the rapid transition into quiescence, focusing on the environmental quenching as its cause, overall the quenching of the whole sample is still dominated by in-situ process related to the supermassive BH, as demonstrated in Fig.~\ref{fig_RF_all}. Particularly for centrals, it is shown that about 80 per cent of the total Gini coefficient is minimized by $M_{\rm BH}$ alone, during the classification process. As mentioned earlier, above BH mass of $10^7\,{\rm M}_\odot$, every galaxy is essentially quenched independent of the other variables. In FLAMINGO, a BH of ${\simeq}\,10^7\,{\rm M}_\odot$ typically resides in about a $10^{12}\,{\rm M}_\odot$ halo, or a $M_\ast\,{\simeq}\,10^{10.5}\,{\rm M}_\odot$ Milky Way-mass galaxy (see the solid lines in Fig.~\ref{fig_fq_ND}), where stellar feedback becomes inefficient and therefore the BHs enter their rapid growth phase \citep{Bower2017, McAlpine2018}. Then as BH masses grow rapidly to $10^8\,{\rm M}_\odot$, which is massive enough to self-regulate the growth via its own AGN feedback, the rapid growth phase ends. During this rapid growth, while BH grows in mass by a factor of 10 or more, the stellar mass of the host galaxy increases from $10^{10.5}\,{\rm M}_\odot$ to $10^{11}\,{\rm M}_\odot$, thus a steep growth in $M_{\rm BH}$--$M_\ast$ plane. Within the stellar mass range between $10^{10}$ and $10^{11}$ solar masses that we probe here, the FLAMINGO simulations are shown to match the relationship between black hole and stellar mass from observations reasonably well when galaxy morphology is taken into account \citep[Fig.~12 of][]{Schaye2023}. Such co-evolution with the rapid BH growth is also consistent with the predictions from EAGLE \citep{Schaye2015, Bower2017}. 

\begin{figure*}
\includegraphics[width=1\linewidth]{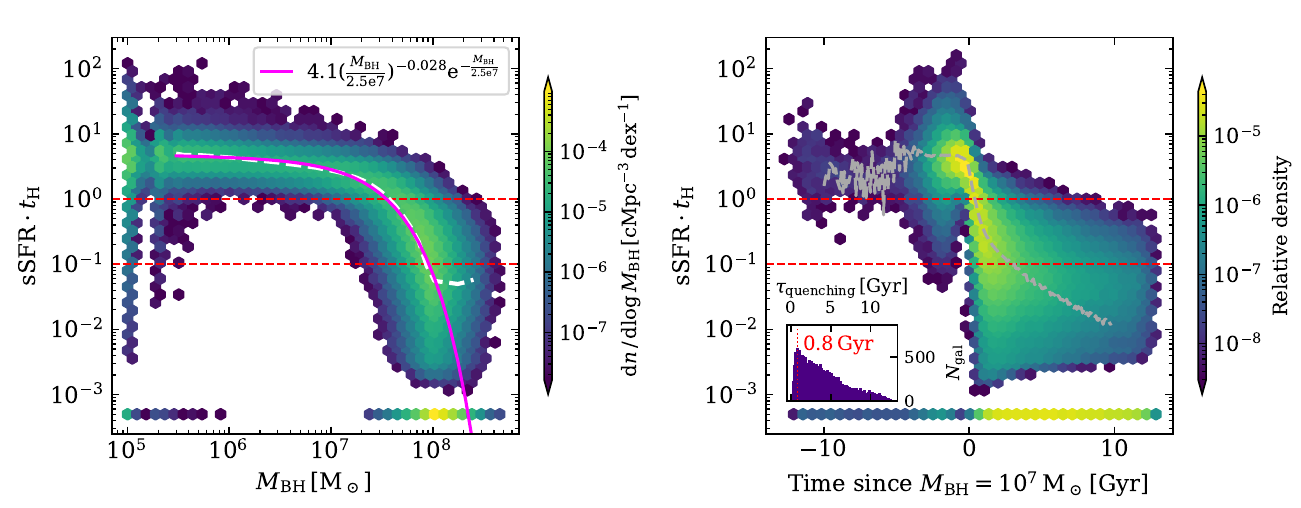}
\caption{({\it Left}): The star-formation history (SFH) of central galaxies in the Main Sample from FLAMINGO as a function of black hole mass, tracked using the simulation merger tree. Those with zero SFR values assigned by the simulation are put near ${\rm sSFR}\cdot t_{\rm H}(z)\,{=}\,5\times10^{-4}$. The horizontal red dashed lines represent ${\rm sSFR}\cdot t_{\rm H}(z)$ of 1 and 0.1, which we define as the `onset' and end of the quenching process, respectively. The white long-dashed line shows the average relation, while a Schechter function fit to it (only fitting for $M_{\rm BH}\,{\leq}\,10^8\,{\rm M}_\odot$) is indicated by the magenta solid curve. ({\it Right}): The SFH as a function of time since a galaxy reaches $M_{\rm BH}\,{=}\,10^7\,{\rm M}_\odot$. Galaxies, on average, are rapidly quenched once passing $M_{\rm BH}\,{=}\,10^7\,{\rm M}_\odot$, with the number of mass $e$-folds falling by more than an order of magnitude over about 1\,Gyr. The inset panel shows the distribution of quenching timescale, $\tau_{\rm quenching}$, defined as time taken for ${\rm sSFR}\cdot t_{\rm H}(z)$ to fall from 1 to 0.1 for each galaxy. The quenching occurs on a broad range of timescales, while peaking around 0.8\,Gyr. }
\label{fig_timescale}
\end{figure*}

While sharp transition of galaxies to quiescence are commonly predicted in many state-of-the-art simulations, the threshold mass and exactly how the transition is driven vary among models. For IllustrisTNG, a rapid increase in the quenched galaxy fraction happens soon after BH mass of $10^8\,{\rm M}_\odot$ \citep[e.g.,][]{Terrazas2020, Piotrowska2022}. This is because the simulation implements the AGN feedback such that, at $M_{\rm BH}\,{\simeq}\,10^8\,{\rm M}_\odot$, BH is assumed to effectively shift from high accretion rate to low accretion-rate mode, and a kinetic jet-like AGN feedback is turned on. The EAGLE simulation has a weaker (less rapid and abrupt) transition into quiescence near a slightly lower BH mass of $10^{7.5}\,{\rm M}_\odot$. Although this lower $M_{\rm BH}$
threshold is closer to observations, which find galaxies are quenched when exceeding $M_{\rm BH}\,{\simeq}\,10^7\,{\rm M}_\odot$, EAGLE fails to quench massive galaxies sufficiently to match the data \citep{Trayford2015, Piotrowska2022}. The SIMBA simulation also implements a transition BH mass of $10^{7.5}\,{\rm M}_\odot$ by `hand', to shift to its jet-mode AGN feedback for galaxies of low Eddington ratio. As a result, BHs in SIMBA stop their rapid growth near the same mass, and begin to quench galaxies effectively \citep{Dave2019, RM2019}. FLAMINGO predicts a sharp and rapid transition to quiescence at $M_{\rm BH}\,{\simeq}\,10^7\,{\rm M}_\odot$, in excellent agreement with observations \citep{Bluck2020b, Chen2020, Piotrowska2022}. This is likely because the accumulated BH energy at $M_{\rm BH}\,{\simeq}\,10^7\,{\rm M}_\odot$ exceeds the binding energy of gas particles within galaxies in FLAMINGO, thus ejecting the gas to the CGM or further. At the same time, we also confirm that the inner CGM gas temperature rapidly rises above $10^5$\,K when $M_{\rm BH}\,{\simeq}\,10^7\,{\rm M}_\odot$ (see Fig.~\ref{fig_fq_Mh}). Stellar and halo mass at the rapid transition to quiescence also range from $M_\ast\,{\simeq}\,10^{10}\,{\rm M}_\odot$ \citep[SIMBA;][]{Dave2019, RM2019} to $M_\ast\,{\simeq}\,10^{11}\,{\rm M}_\odot$ or $M_{\rm h}\,{\simeq}\,10^{13}\,{\rm M}_\odot$ \citep[EAGLE; e.g.,][]{Piotrowska2022}. The prediction by FLAMINGO that the quenching becomes effective near the Milky Way mass lies in between. 

Observations and theoretical models suggest a broad and/or bimodal distribution of quenching timescales at $z\,{\lesssim}\,2$ \citep[e.g.,][]{Wu2018, Belli2019, RM2019, Wild2020, Park2022, Tacchella2022}. As discussed in earlier sections, our findings in Figs.~\ref{fig_sSFHfqz}--\ref{fig_fq_Mh} imply that the quenching takes place on a short timescale of $\tau_{\rm quenching}\,{\lesssim}\,1$\,Gyr, thus supporting ``fast quenching", mainly driven by the AGN feedback heating and ejecting the gas around galaxies (Fig.~\ref{fig_fq_Mh}). This is also in a broad agreement with the previous findings from the EAGLE simulation \citep[e.g.,][]{Trayford2016, Correa2019}. Here we directly probe the distribution of quenching timescales in relation to BH mass, utilizing the simulation merger tree. We define the quenching timescale, $\tau_{\rm quenching}$, as the time interval for ${\rm sSFR}\cdot t_{\rm H}(z)$ to drop from 1 to 0.1. As mentioned earlier, this is roughly a conventional average for defining quenching in the literature \citep[e.g.,][]{Gallazzi2014, Pacifici2016}. We further define the onset of quenching as when the number of mass $e$-folds falls below 1 for the last time in a galaxy's history. We checked that using more sophisticated criteria, such as running averages to avoid an instantaneous drop in SFR being misidentified as quenching, does not change our conclusions. We also count only the latest quenching events, namely neglecting earlier quenchings in case of rejuvenations and multiple quenchings, which we do not find to be very common in FLAMINGO, in agreement with other studies and observations \citep[e.g.,][]{RM2019, Tacchella2022}. Fig.~\ref{fig_timescale}, in the left panel, shows the number of mass $e$-folds for central galaxies in Main Sample, as a function of BH mass. A Schechter function fit (magenta) to the average (white) for $M_{\rm BH}\,{\leq}\,10^8\,{\rm M}_\odot$ finds a characteristic mass of $M_{\rm BH}\,{\simeq}\,2.5\times 10^7\,{\rm M}_\odot$, which also coincides roughly with BH mass at the onset of quenching, $M_{\rm BH,onset}$ where the number of mass $e$-folds falls below 1. Note that this is a slightly higher value than $10^7\,{\rm M}_\odot$ that we found earlier as a threshold BH mass for quenching at $z\,{=}\,0$. This is because the result here contains galaxies of various quenching epochs, with a majority of them quenched even as early as at $z\,{\gtrsim}\,4$, and the threshold BH mass varies with time, typically increasing with redshift due to accelerated cooling (see Fig.~\ref{fig_fq_ND_z1} and the discussion thereof). The right panel presents the star-formation history as a function of time, using $M_{\rm BH}\,{=}\,10^7\,{\rm M}_\odot$ as a pivot mass. It is indeed found that the sSFR rapidly decreases once reaching $M_{\rm BH}\,{=}\,10^7\,{\rm M}_\odot$, dropping by more than an order of magnitude in about a Gyr timespan. This confirms our earlier conjecture that galaxies are quenched rapidly around $M_{\rm BH}\,{=}\,10^7\,{\rm M}_\odot$. However, the average curve (grey) also indicates a wide dispersion of quenching timescale, particularly changing slopes in the middle as in a broken double power-law, which hints a bimodal distribution with two populations of fast and slow quenching. Note that galaxies located on the side of slower quenching, are those quenched earlier and having reached the threshold BH mass at higher redshifts. They are typically massive in stellar and halo mass as well. This means that more massive systems reach a threshold BH mass earlier, form earlier, and then quench slowly, while less massive ones form slowly to reach the onset BH mass just lately, and then became inactive recently and rapidly. This is consistent with \citet{RM2019}, who used the SIMBA simulation to find a fast-mode quenching for galaxies of $M_\ast\,{\simeq}\,10^{10}$--$10^{10.5}\,{\rm M}_\odot$, while slower quenching dominates at higher mass. The inset panel shows the distribution of the quenching timescale, demonstrating that the galaxy quenching occurs on a variety of timescales, while peaking around 0.8\,Gyr. Our results are slightly sensitive to the choice of definition for quenching. For example, using higher cuts for the onset and quenching is found to reveal a clearer bimodal distribution with a sharp concentration typically around 0.6\,Gyr and a broad distribution for the rest. This prediction is in a broad agreement with observational studies such as \cite{Tacchella2022}, who found about 25 per cent of quiescent galaxies at $z\,{\simeq}\,0.8$ were quenched within 500\,Myr, with an overall median of $\tau_{\rm quenching}\,{=}\,$0.1--1.8\,Gyr (see also \citealt{Park2024}).

We also investigated the correlations of each parameter in the feature set, in a traditional way, with the quenched fraction of galaxies. For centrals, we find that only the BH mass is positively correlated with the quenched fraction, as discussed above. An overall negative correlation with halo mass is found, while there exists a marginal positive correlation after the turn-over around halo mass of $10^{12}\,{\rm M}_\odot$ which, however, is diluted in the overall relation when weighted by the numbers. The average trend is dominated by the star formation being most efficient within the Milky Way-mass halo and falling in the lower mass. The halo concentration, $R_{200}/R_{2500}$ used as a proxy, is also found to present a negative correlation with the quenched fraction. This is opposite to the expectation, as it is well known that the concentration is a strong function of halo mass, with more massive haloes being less concentrated. Thus given the negative dependence of quenching on halo mass, a positive correlation with the concentration parameter is expected. The negative correlation, instead, may reflect a higher binding energy in haloes of a higher concentration at a given halo mass, which holds the gas against being ejected. However, the strength of its impact as found from the RF analysis is only marginal, as seen in Fig.~\ref{fig_RF_all}, in which case interpretation of the feature is highly cautioned in general.

\subsection{The main drivers for quenching satellites}
\label{ssec_disc_sat}

We demonstrated, in Fig.~\ref{fig_RF_all}, that the black hole mass is also the most direct in-situ parameter for quiescence of satellites. The same characteristic mass of $M_{\rm BH}\,{\simeq}\,10^7\,{\rm M}_\odot$, and the rapid transition from the main sequence to quiescence across the mass, are found for satellites similarly to centrals. Compared to the centrals, however, the relative importance given to $M_{\rm BH}$ in quenching is reduced. While part of this is attributed to the more features included for satellites over which the total relative importance is shared, it is notable that stellar mass is given a remarkable importance unlike for centrals. This can be understood in the context of pre-quenching, namely quenching that occurs prior to the galaxies' infall to the current host halo. Before merging into the current environment, satellites have been either centrals or in another dense environment. As there is no `current' property in the feature set that represents the condition of their previous environment, it may be understood that the history of past environments are reflected in the apparent significance given to stellar mass. But which one, then, is the more significant driver for pre-quenching satellites: in-situ quenching from when they were centrals, or ex-situ suppression in the previous host halo?

Fortunately, there seems to be a way to answer this question, thanks to the opposite signs of correlations between the two scenarios. As seen in Figs.~\ref{fig_sSFHfqz} and \ref{fig_fq_ND}, and discussed earlier, the environmental quenching indicates that satellites in more massive haloes are more frequently found quenched, namely a positive correlation with the quenched fraction. The in-situ quenching related to halo mass, on the other hand, is found to be in a negative correlation, as shown for centrals where those in more massive haloes are less quenched at a given BH and stellar mass. Therefore by investigating the correlation of stellar mass with quiescent satellites (assuming stellar mass is a proxy of the previous halo mass), one can determine which process is the more dominant. On average, we find that the quenched fraction is found to be higher among lower-mass satellites. We thus conclude that the quenching mechanism found for satellites indirectly via their stellar mass is due to the shallower potential well that failed to hold gas against stellar feedback in the past, prior to their infall to the current host halo. 

Interestingly, there is a marginal hint that the supermassive BH of the central galaxy in the halo ($M_{\rm BH,host}$) also affects the quenching of satellites. We find a positive correlation that satellites with a more massive central BH (i.e. black hole in their host halo, not one in a satellite itself) are found to be more quenched. This may indicate that the energy injection from the central AGN feedback is powerful enough to even heat or blow out gas surrounding satellites to further suppress their star formation. However, it is also rather interesting to note how small the importance given to $M_{\rm BH, host}$ is relative to that to $M_{\rm BH}$, the satellite's own BH mass. This may imply that the quenching effect of AGN feedback is an accumulated result from its multiple events over the history, rather than an impact from an instantaneous state at any given time. Satellites, therefore, which are normally found with not much of their lifetime spent inside the present local environment, would not have had enough time to be affected and quenched by the central BH of the current host halo.

\begin{figure}
\includegraphics[width=1.01\linewidth]{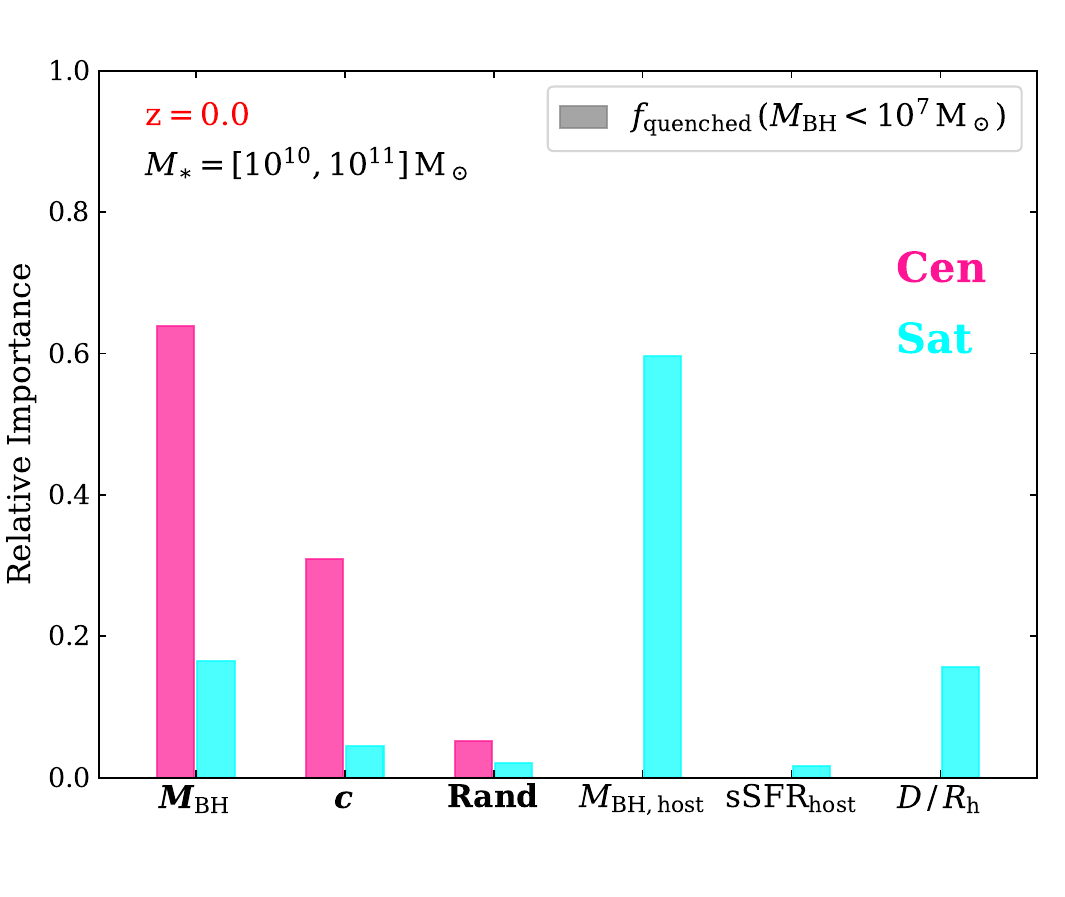}
\vspace{-0.8cm}
\caption{Relative importance of in-situ and ex-situ drivers for quenching, determined from the Random Forest (RF) analysis for Main Sample-lowBH from the $z\,{=}\,0$ snapshot of FLAMINGO. This is the same as Fig.~\ref{fig_RF_all} but for the reduced feature set in order to explore impacts in a case that the true quenching drivers are not included. The reduced set includes black hole (BH) mass ($M_{\rm BH}$), halo concentration proxy of $R_{200}/R_{2500}$ ($c$), random number (Rand), BH mass of host halo ($M_{\rm BH, host}$), specific star-formation rate of central galaxy (${\rm sSFR_{host}}$), and distance from the center of host halo divided by halo radius ($D/R_{\rm h}$). Compared with Fig.~\ref{fig_RF_all}, the halo concentration and the host BH mass are given highly enhanced importance for centrals and satellites, respectively, potentially misleading interpretations. The features for centrals are in boldface. }
\label{fig_RF_reduced}
\end{figure}

\subsection{Limitations of the Random Forest analysis}

As briefly touched upon in, e.g., Sect.~\ref{ssec_drivers}, there are several limitations in the RF analysis, which are general issues in approaches using the classifier. First of all, although better than traditional correlation analysis, RF itself explores correlations, and its potential to identify causation still depends on interpretations based on a priori knowledge. Second, in order to reveal a parameter truly responsible for quenching, the parameter must be identified and included in the input set of features. Otherwise the power of the RF to distinguish between direct and nuisance variables is limited, still potentially mis-identifying a highly-correlated secondary parameter as a direct cause. Fig.~\ref{fig_RF_reduced} demonstrates such effect in an extreme test case where stellar mass, halo mass, and $M_{\rm \ast, 1.5pMpc}$ are removed from the input feature set into the RF for Main Sample-lowBH. The methodology of the analysis is kept the same otherwise as in Fig.~\ref{fig_RF_all} and as described in Sect.~\ref{ssec_RF}. Notably, now with the features of the most significance for quenching removed for both centrals and satellites, the halo concentration (for centrals) and $M_{\rm BH, host}$ (for satellites) have their relative importance substantially boosted. Particularly, for satellites, now the host halo's BH mass suddenly appears as the most dominant property for quenching, being even more impactful than its own BH mass, which we know is not true from the previous result. Also, the discussion in Sect.~\ref{ssec_disc_sat} on the stellar-mass dependence of satellite quenching is another example that complicates an interpretation because of a lack of more direct features to prove or disprove particular scenarios in consideration. However, one cannot expand the input set to include every parameter potentially relevant for quenching mechanisms, since increasing the number of variables will have the relative importance shared and reduced, making it difficult to identify which feature is genuinely responsible. This is particularly true when the measurements and relations between variables are noisy, as is common in astronomical data, and even more true if some of the variables are highly correlated. This limitation is well demonstrated by B22, who found that nuisance parameters are given non-negligible importance scores when noise is added, making naive interpretation dangerous (see also \citealt{Goubert2024}).

\section[summary]{Summary}
\label{sec_summary}

In this paper, we used the highest-resolution (`L1\_m8') FLAMINGO simulation to explore the characteristics of quiescent and star-forming main-sequence galaxies, by constructing a large sample of 5.3 million galaxies with mass similar to the Milky Way, taking advantage of the large box size of FLAMINGO. We restricted our analysis carefully to the mass range of $M_\ast\,{=}\,10^{10}-10^{11}\,{\rm M}_\odot$, in order to avoid the limitation of the numerical resolution as well as uncertainties in the model prediction for the most massive galaxies. We define galaxies with ${\rm sSFR}\cdot t_{\rm H}(z)<0.1$ ($>0.1$) as quiescent (star-forming), where sSFR is the instantaneous sSFR. However, we find our analysis and conclusions are robust against a different choice of the threshold, by up to a factor of 5. The simulation reproduces the overall evolutionary trends of the stellar mass function (SMF) observed in the data, including the increase in passive fraction with redshift, as well as the mass scale at which passive galaxies dominate (within a factor of about 2). However, the simulated quiescent fraction is systematically lower than observed, though differences in selection criteria complicate a direct comparison. We employed Random Forest (RF) classifiers to identify and discern the properties of galaxies and environments that are more directly related to quenching, from secondary, indirectly correlated parameters. Our findings are summarized as follows. 

First, in Sect.~\ref{ssec_MSQ} and Fig.~\ref{fig_sSFHfqz}, we have shown that the main-sequence galaxies present strikingly similar sSFR regardless of their environment or local dominance (namely whether a galaxy is a central or satellite), whereas the quenched fraction is highly sensitive to both the environment and local dominance. This implies a rapid transition from the main sequence to quiescence, and that whatever physical mechanisms are responsible for the quenching must act effectively on a relatively short timescale, as we directly demonstrated in Sect.~\ref{ssec_disc_cen} and Fig.~\ref{fig_timescale}. These findings and conclusions hold from $z\,{\simeq}\,7$ to $z\,{=}\,0$, i.e. for the whole cosmic history probed in our analysis. 

Second, in Sect.~\ref{ssec_drivers}, and in Figs.~\ref{fig_fq_ND} and \ref{fig_fq_Mh}, we identified that there exist two distinct regimes for galaxy quenching: at $M_{\rm BH}\,{\gtrsim}\,10^7\,{\rm M}_\odot$, essentially all galaxies are predicted to be quenched, whereas at $M_{\rm BH}\,{\lesssim}\,10^7\,{\rm M}_\odot$ quenching is correlated with halo mass and local stellar mass, both of which are proxies of environment, with the quiescent fraction found to be minimum, about 0.2, in $M_{\rm h}\,{\simeq}\,10^{12}\,{\rm M}_\odot$ haloes, above which the fraction increases rapidly. The transition from the main sequence to quiescence takes place rapidly around $M_{\rm BH}\,{=}\,10^7\,{\rm M}_\odot$. The characteristic BH mass is insensitive to the details of the definitions of quiescent and main-sequence galaxies, because of the sharp transition that occurs within a narrow BH mass range. This BH mass of $10^7\,{\rm M}_\odot$ being critical for galaxy quenching, as well as the rapid BH growth near the mass, are in excellent agreement with observations \citep[e.g.,][]{Terrazas2016, Piotrowska2022}. The rapid transition to quiescence near this BH mass is thought to be due to accumulated BH energy exceeding the binding energy \citep[e.g.,][]{Davies2019, Davies2020, Oppenheimer2020} or thermal energy \citep[e.g.,][]{Zinger2020} required for gas to be ejected or heated and cease star formation. Indeed, in the bottom panels of Fig.~\ref{fig_fq_Mh}, we investigated the gas properties of the inner CGM (within 50\,pkpc), and found that the gas mass fraction (with respect to halo mass), star-formation efficiency (SFE), and gas temperature all sharply change near $M_{\rm BH}\,{=}\,10^7\,{\rm M}_\odot$, approximately by a factor of 3, 10, and 100, respectively. Both the SFE and gas mass fraction trace the inner CGM amount (with respect to the ISM and halo mass, respectively). The gas properties of the outer CGM present much weaker transitions near $M_{\rm BH}\,{=}\,10^7\,{\rm M}_\odot$. The results indicate that galaxies are quenched by the AGN feedback via the inner CGM fraction, which can be affected by ejection of the ISM into the inner CGM, by ejection of the inner CGM into the outer region, and by preventive impact on the CGM to accrete on to galaxies. All these findings demonstrate that AGN feedback is the most dominant physical process for quenching massive galaxies of $M_\ast\,{\gtrsim}\,10^{10.5}\,{\rm M}_\odot$, regardless of whether a galaxy is a central or a satellite, in any environment.  

Third, in Fig.~\ref{fig_RF_all}, motivated by the finding of the critical BH mass, we carried out a RF analysis to identify the physical properties directly related to the environmental quenching of galaxies with $M_{\rm BH}\,{\lesssim}\,10^7\,{\rm M}_\odot$. Among a set of environmental proxies, including the total stellar mass within 1.5\,pMpc, halo concentration, sSFR and BH mass of host halo, and distance to host halo center, the RF identifies halo mass as the most direct parameter related to quenching the low-mass galaxies below $M_{\rm BH}\,{\simeq}\,10^7\,{\rm M}_\odot$, where the importance of BH mass is found to be substantially reduced but still the second most impactful parameter. A short timescale inferred by the rapid transition, combined with the similar trends found over a range of mass and redshift, hints that the so-called galaxy harassment, rather than tidal or ram-pressure stripping, is likely the dominant mechanism for quenching the massive satellites. Also, a notable importance given by RF to stellar mass of satellites implies in-situ pre-quenching rather than ex-situ preprocessing in their previous host, prior to their infall to the current host. 

The above conclusions regarding the two distinct evolution regimes, the rapid transition between them across the critical BH mass of $10^7\,{\rm M}_\odot$, and the dominance of BH and halo mass as the most direct properties for the quenching, are found to hold true at higher redshifts in the cosmic afternoon of up to $z\,{\simeq}\,1$, as shown in Figs.~\ref{fig_fq_ND_z1} and \ref{fig_RF_all_z1}.  

Finally, in Fig.~\ref{fig_timescale}, we directly probed the star-formation history (SFH) as a function of BH mass, and the timescale of quenching. A Schechter function fits the SFH reasonably well, with a best-fit pivot BH mass of about 2$\times10^7\,{\rm M_\odot}$, similar to the characteristic mass identified in Figs.~\ref{fig_fq_ND} and \ref{fig_fq_Mh}. The quenching timescale is found to peak around 0.8\,Gyr, indicating fast quenching, while the whole distribution is broad, likely due to various quenching mechanisms such as those driven by halo and environments, and due to evolution of the processes and their timescales. 

In this work, we examined the galaxy properties and environment proxies related to the quenching of Milky Way-mass galaxies in the cosmic afternoon, using the large-volume FLAMINGO simulation also aided by RF classifiers. The dynamic range of samples in mass, retaining a large sample size, may be expanded with future simulations where the numerical resolution can be improved with accelerated computing power, which will help analyze the quenching mechanisms in smaller or more massive haloes. Also, more detailed and robust attempts to identify the characteristic BH mass in observational data are encouraged. This would shed light on the growth of supermassive BHs and the timescale over which AGN feedback impacts galaxies in the galaxy--BH co-evolution, arguably the most unfilled region in the modern picture of galaxy formation theory. 



\section*{ACKNOWLEDGEMENTS}

We thank the referee for constructive comments that improved the paper. SL and RM acknowledge support by the Science and Technology Facilities Council (STFC) and by the UKRI Frontier Research grant RISEandFALL. RM also acknowledges funding from a research professorship from the Royal Society. This work used the DiRAC@Durham facility managed by the Institute for Computational Cosmology on behalf of the STFC DiRAC HPC Facility (\url{www.dirac.ac.uk}). The equipment was funded by BEIS capital funding via STFC capital grants ST/K00042X/1, ST/P002293/1, ST/R002371/1 and ST/S002502/1, Durham University and STFC operations grant ST/R000832/1. DiRAC is part of the National e-Infrastructure. This work is partly funded by research programme Athena 184.034.002 from the Dutch Research Council (NWO).

\section*{DATA AVAILABILITY}

The data underlying this article will be shared on reasonable request to the corresponding author.

\bibliographystyle{mnras}
\bibliography{GalEnv.bib}

\begin{thebibliography}{}
\makeatletter
\relax
\def\mn@urlcharsother{\let\do\@makeother \do\$\do\&\do\#\do\^\do\_\do\%\do\~}
\def\mn@doi{\begingroup\mn@urlcharsother \@ifnextchar [ {\mn@doi@}
  {\mn@doi@[]}}
\def\mn@doi@[#1]#2{\def\@tempa{#1}\ifx\@tempa\@empty \href
  {http://dx.doi.org/#2} {doi:#2}\else \href {http://dx.doi.org/#2} {#1}\fi
  \endgroup}
\def\mn@eprint#1#2{\mn@eprint@#1:#2::\@nil}
\def\mn@eprint@arXiv#1{\href {http://arxiv.org/abs/#1} {{\tt arXiv:#1}}}
\def\mn@eprint@dblp#1{\href {http://dblp.uni-trier.de/rec/bibtex/#1.xml}
  {dblp:#1}}
\def\mn@eprint@#1:#2:#3:#4\@nil{\def\@tempa {#1}\def\@tempb {#2}\def\@tempc
  {#3}\ifx \@tempc \@empty \let \@tempc \@tempb \let \@tempb \@tempa \fi \ifx
  \@tempb \@empty \def\@tempb {arXiv}\fi \@ifundefined
  {mn@eprint@\@tempb}{\@tempb:\@tempc}{\expandafter \expandafter \csname
  mn@eprint@\@tempb\endcsname \expandafter{\@tempc}}}

\bibitem[\protect\citeauthoryear{{Abadi}, {Moore}  \& {Bower}}{{Abadi}
  et~al.}{1999}]{AbadiMooreBower1999}
{Abadi} M.~G.,  {Moore} B.,   {Bower} R.~G.,  1999, \mn@doi [\mnras]
  {10.1046/j.1365-8711.1999.02715.x}, \href
  {https://ui.adsabs.harvard.edu/abs/1999MNRAS.308..947A} {308, 947}

\bibitem[\protect\citeauthoryear{{Abbott} et~al.,}{{Abbott}
  et~al.}{2022}]{Abbott2022}
{Abbott} T.~M.~C.,  et~al., 2022, \mn@doi [\prd] {10.1103/PhysRevD.105.023520},
  \href {https://ui.adsabs.harvard.edu/abs/2022PhRvD.105b3520A} {105, 023520}

\bibitem[\protect\citeauthoryear{{Bah{\'e}} et~al.,}{{Bah{\'e}}
  et~al.}{2022}]{Bahe2022}
{Bah{\'e}} Y.~M.,  et~al., 2022, \mn@doi [\mnras] {10.1093/mnras/stac1339},
  \href {https://ui.adsabs.harvard.edu/abs/2022MNRAS.516..167B} {516, 167}

\bibitem[\protect\citeauthoryear{{Baker} et~al.,}{{Baker}
  et~al.}{2024a}]{Baker2025}
{Baker} W.~M.,  et~al., 2024a, \mn@doi [arXiv e-prints]
  {10.48550/arXiv.2410.14773}, \href
  {https://ui.adsabs.harvard.edu/abs/2024arXiv241014773B} {p. arXiv:2410.14773}

\bibitem[\protect\citeauthoryear{{Baker} et~al.,}{{Baker}
  et~al.}{2024b}]{Baker2024}
{Baker} W.~M.,  et~al., 2024b, \mn@doi [\mnras] {10.1093/mnras/stae2059}, \href
  {https://ui.adsabs.harvard.edu/abs/2024MNRAS.534...30B} {534, 30}

\bibitem[\protect\citeauthoryear{{Balogh} \& {Morris}}{{Balogh} \&
  {Morris}}{2000}]{BaloghMorris2000}
{Balogh} M.~L.,  {Morris} S.~L.,  2000, \mn@doi [\mnras]
  {10.1046/j.1365-8711.2000.03826.x}, \href
  {https://ui.adsabs.harvard.edu/abs/2000MNRAS.318..703B} {318, 703}

\bibitem[\protect\citeauthoryear{{Balogh}, {Pearce}, {Bower}  \&
  {Kay}}{{Balogh} et~al.}{2001}]{Balogh2001}
{Balogh} M.~L.,  {Pearce} F.~R.,  {Bower} R.~G.,   {Kay} S.~T.,  2001, \mn@doi
  [\mnras] {10.1111/j.1365-2966.2001.04667.x}, \href
  {https://ui.adsabs.harvard.edu/abs/2001MNRAS.326.1228B} {326, 1228}

\bibitem[\protect\citeauthoryear{{Balogh}, {Baldry}, {Nichol}, {Miller},
  {Bower}  \& {Glazebrook}}{{Balogh} et~al.}{2004}]{Balogh2004}
{Balogh} M.~L.,  {Baldry} I.~K.,  {Nichol} R.,  {Miller} C.,  {Bower} R.,
  {Glazebrook} K.,  2004, \mn@doi [\apjl] {10.1086/426079}, \href
  {https://ui.adsabs.harvard.edu/abs/2004ApJ...615L.101B} {615, L101}

\bibitem[\protect\citeauthoryear{{Balogh} et~al.,}{{Balogh}
  et~al.}{2016}]{Balogh2016}
{Balogh} M.~L.,  et~al., 2016, \mn@doi [\mnras] {10.1093/mnras/stv2949}, \href
  {https://ui.adsabs.harvard.edu/abs/2016MNRAS.456.4364B} {456, 4364}

\bibitem[\protect\citeauthoryear{{Behroozi}, {Wechsler}  \&
  {Conroy}}{{Behroozi} et~al.}{2013}]{Behroozi2013}
{Behroozi} P.~S.,  {Wechsler} R.~H.,   {Conroy} C.,  2013, \mn@doi [\apj]
  {10.1088/0004-637X/770/1/57}, \href
  {https://ui.adsabs.harvard.edu/abs/2013ApJ...770...57B} {770, 57}

\bibitem[\protect\citeauthoryear{{Belli}, {Newman}  \& {Ellis}}{{Belli}
  et~al.}{2019}]{Belli2019}
{Belli} S.,  {Newman} A.~B.,   {Ellis} R.~S.,  2019, \mn@doi [\apj]
  {10.3847/1538-4357/ab07af}, \href
  {https://ui.adsabs.harvard.edu/abs/2019ApJ...874...17B} {874, 17}

\bibitem[\protect\citeauthoryear{{Best} \& {Heckman}}{{Best} \&
  {Heckman}}{2012}]{BestHeckman2012}
{Best} P.~N.,  {Heckman} T.~M.,  2012, \mn@doi [\mnras]
  {10.1111/j.1365-2966.2012.20414.x}, \href
  {https://ui.adsabs.harvard.edu/abs/2012MNRAS.421.1569B} {421, 1569}

\bibitem[\protect\citeauthoryear{{Binney}}{{Binney}}{2004}]{Binney2004}
{Binney} J.,  2004, \mn@doi [\mnras] {10.1111/j.1365-2966.2004.07277.x}, \href
  {https://ui.adsabs.harvard.edu/abs/2004MNRAS.347.1093B} {347, 1093}

\bibitem[\protect\citeauthoryear{{Blandford}, {Meier}  \&
  {Readhead}}{{Blandford} et~al.}{2019}]{BlandfordMeierReadhead2019}
{Blandford} R.,  {Meier} D.,   {Readhead} A.,  2019, \mn@doi [\araa]
  {10.1146/annurev-astro-081817-051948}, \href
  {https://ui.adsabs.harvard.edu/abs/2019ARA&A..57..467B} {57, 467}

\bibitem[\protect\citeauthoryear{{Bluck}, {Maiolino}, {S{\'a}nchez}, {Ellison},
  {Thorp}, {Piotrowska}, {Teimoorinia}  \& {Bundy}}{{Bluck}
  et~al.}{2020a}]{Bluck2020a}
{Bluck} A. F.~L.,  {Maiolino} R.,  {S{\'a}nchez} S.~F.,  {Ellison} S.~L.,
  {Thorp} M.~D.,  {Piotrowska} J.~M.,  {Teimoorinia} H.,   {Bundy} K.~A.,
  2020a, \mn@doi [\mnras] {10.1093/mnras/stz3264}, \href
  {https://ui.adsabs.harvard.edu/abs/2020MNRAS.492...96B} {492, 96}

\bibitem[\protect\citeauthoryear{{Bluck} et~al.,}{{Bluck}
  et~al.}{2020b}]{Bluck2020b}
{Bluck} A. F.~L.,  et~al., 2020b, \mn@doi [\mnras] {10.1093/mnras/staa2806},
  \href {https://ui.adsabs.harvard.edu/abs/2020MNRAS.499..230B} {499, 230}

\bibitem[\protect\citeauthoryear{{Bluck}, {Maiolino}, {Brownson}, {Conselice},
  {Ellison}, {Piotrowska}  \& {Thorp}}{{Bluck} et~al.}{2022}]{Bluck2022}
{Bluck} A. F.~L.,  {Maiolino} R.,  {Brownson} S.,  {Conselice} C.~J.,
  {Ellison} S.~L.,  {Piotrowska} J.~M.,   {Thorp} M.~D.,  2022, \mn@doi [\aap]
  {10.1051/0004-6361/202142643}, \href
  {https://ui.adsabs.harvard.edu/abs/2022A&A...659A.160B} {659, A160}

\bibitem[\protect\citeauthoryear{{Bluck}, {Piotrowska}  \& {Maiolino}}{{Bluck}
  et~al.}{2023}]{Bluck2023}
{Bluck} A. F.~L.,  {Piotrowska} J.~M.,   {Maiolino} R.,  2023, \mn@doi [\apj]
  {10.3847/1538-4357/acac7c}, \href
  {https://ui.adsabs.harvard.edu/abs/2023ApJ...944..108B} {944, 108}

\bibitem[\protect\citeauthoryear{{Bluck} et~al.,}{{Bluck}
  et~al.}{2024}]{Bluck2024}
{Bluck} A. F.~L.,  et~al., 2024, \mn@doi [\apj] {10.3847/1538-4357/ad0a98},
  \href {https://ui.adsabs.harvard.edu/abs/2024ApJ...961..163B} {961, 163}

\bibitem[\protect\citeauthoryear{{Booth} \& {Schaye}}{{Booth} \&
  {Schaye}}{2009}]{BoothSchaye2009}
{Booth} C.~M.,  {Schaye} J.,  2009, \mn@doi [\mnras]
  {10.1111/j.1365-2966.2009.15043.x}, \href
  {https://ui.adsabs.harvard.edu/abs/2009MNRAS.398...53B} {398, 53}

\bibitem[\protect\citeauthoryear{{Booth} \& {Schaye}}{{Booth} \&
  {Schaye}}{2011}]{BoothSchaye2011}
{Booth} C.~M.,  {Schaye} J.,  2011, \mn@doi [\mnras]
  {10.1111/j.1365-2966.2011.18203.x}, \href
  {https://ui.adsabs.harvard.edu/abs/2011MNRAS.413.1158B} {413, 1158}

\bibitem[\protect\citeauthoryear{{Boselli} \& {Gavazzi}}{{Boselli} \&
  {Gavazzi}}{2006}]{BoselliGavazzi2006}
{Boselli} A.,  {Gavazzi} G.,  2006, \mn@doi [\pasp] {10.1086/500691}, \href
  {https://ui.adsabs.harvard.edu/abs/2006PASP..118..517B} {118, 517}

\bibitem[\protect\citeauthoryear{{Bower}, {Benson}, {Malbon}, {Helly}, {Frenk},
  {Baugh}, {Cole}  \& {Lacey}}{{Bower} et~al.}{2006}]{Bower2006}
{Bower} R.~G.,  {Benson} A.~J.,  {Malbon} R.,  {Helly} J.~C.,  {Frenk} C.~S.,
  {Baugh} C.~M.,  {Cole} S.,   {Lacey} C.~G.,  2006, \mn@doi [\mnras]
  {10.1111/j.1365-2966.2006.10519.x}, \href
  {https://ui.adsabs.harvard.edu/abs/2006MNRAS.370..645B} {370, 645}

\bibitem[\protect\citeauthoryear{{Bower}, {McCarthy}  \& {Benson}}{{Bower}
  et~al.}{2008}]{Bower2008}
{Bower} R.~G.,  {McCarthy} I.~G.,   {Benson} A.~J.,  2008, \mn@doi [\mnras]
  {10.1111/j.1365-2966.2008.13869.x}, \href
  {https://ui.adsabs.harvard.edu/abs/2008MNRAS.390.1399B} {390, 1399}

\bibitem[\protect\citeauthoryear{{Bower}, {Schaye}, {Frenk}, {Theuns},
  {Schaller}, {Crain}  \& {McAlpine}}{{Bower} et~al.}{2017}]{Bower2017}
{Bower} R.~G.,  {Schaye} J.,  {Frenk} C.~S.,  {Theuns} T.,  {Schaller} M.,
  {Crain} R.~A.,   {McAlpine} S.,  2017, \mn@doi [\mnras]
  {10.1093/mnras/stw2735}, \href
  {https://ui.adsabs.harvard.edu/abs/2017MNRAS.465...32B} {465, 32}

\bibitem[\protect\citeauthoryear{{Carnall} et~al.,}{{Carnall}
  et~al.}{2023}]{Carnall2023}
{Carnall} A.~C.,  et~al., 2023, \mn@doi [\mnras] {10.1093/mnras/stad369}, \href
  {https://ui.adsabs.harvard.edu/abs/2023MNRAS.520.3974C} {520, 3974}

\bibitem[\protect\citeauthoryear{{Carnall} et~al.,}{{Carnall}
  et~al.}{2024}]{Carnall2024}
{Carnall} A.~C.,  et~al., 2024, \mn@doi [\mnras] {10.1093/mnras/stae2092},
  \href {https://ui.adsabs.harvard.edu/abs/2024MNRAS.534..325C} {534, 325}

\bibitem[\protect\citeauthoryear{{Chabrier}}{{Chabrier}}{2003}]{Chabrier2003}
{Chabrier} G.,  2003, \mn@doi [\pasp] {10.1086/376392}, \href
  {https://ui.adsabs.harvard.edu/abs/2003PASP..115..763C} {115, 763}

\bibitem[\protect\citeauthoryear{{Chaikin}, {Schaye}, {Schaller},
  {Ben{\'\i}tez-Llambay}, {Nobels}  \& {Ploeckinger}}{{Chaikin}
  et~al.}{2023}]{Chaikin2023}
{Chaikin} E.,  {Schaye} J.,  {Schaller} M.,  {Ben{\'\i}tez-Llambay} A.,
  {Nobels} F. S.~J.,   {Ploeckinger} S.,  2023, \mn@doi [\mnras]
  {10.1093/mnras/stad1626}, \href
  {https://ui.adsabs.harvard.edu/abs/2023MNRAS.523.3709C} {523, 3709}

\bibitem[\protect\citeauthoryear{Chawla, Bowyer, Hall  \& Kegelmeyer}{Chawla
  et~al.}{2002}]{Chawla2002}
Chawla N.,  Bowyer K.,  Hall L.,   Kegelmeyer W.,  2002, \mn@doi [J. Artif.
  Intell. Res. (JAIR)] {10.1613/jair.953}, 16, 321

\bibitem[\protect\citeauthoryear{Chen et~al.,}{Chen et~al.}{2020}]{Chen2020}
Chen Z.,  et~al., 2020, \mn@doi [The Astrophysical Journal]
  {10.3847/1538-4357/ab9633}, 897, 102

\bibitem[\protect\citeauthoryear{{Choi}, {Ostriker}, {Naab}  \&
  {Johansson}}{{Choi} et~al.}{2012}]{Choi2012}
{Choi} E.,  {Ostriker} J.~P.,  {Naab} T.,   {Johansson} P.~H.,  2012, \mn@doi
  [\apj] {10.1088/0004-637X/754/2/125}, \href
  {https://ui.adsabs.harvard.edu/abs/2012ApJ...754..125C} {754, 125}

\bibitem[\protect\citeauthoryear{{Cole}, {Aragon-Salamanca}, {Frenk}, {Navarro}
   \& {Zepf}}{{Cole} et~al.}{1994}]{Cole1994}
{Cole} S.,  {Aragon-Salamanca} A.,  {Frenk} C.~S.,  {Navarro} J.~F.,   {Zepf}
  S.~E.,  1994, \mn@doi [\mnras] {10.1093/mnras/271.4.781}, \href
  {https://ui.adsabs.harvard.edu/abs/1994MNRAS.271..781C} {271, 781}

\bibitem[\protect\citeauthoryear{{Cole}, {Lacey}, {Baugh}  \& {Frenk}}{{Cole}
  et~al.}{2000}]{Cole2000}
{Cole} S.,  {Lacey} C.~G.,  {Baugh} C.~M.,   {Frenk} C.~S.,  2000, \mn@doi
  [\mnras] {10.1046/j.1365-8711.2000.03879.x}, \href
  {https://ui.adsabs.harvard.edu/abs/2000MNRAS.319..168C} {319, 168}

\bibitem[\protect\citeauthoryear{{Colombo} et~al.,}{{Colombo}
  et~al.}{2020}]{Colombo2020}
{Colombo} D.,  et~al., 2020, \mn@doi [\aap] {10.1051/0004-6361/202039005},
  \href {https://ui.adsabs.harvard.edu/abs/2020A&A...644A..97C} {644, A97}

\bibitem[\protect\citeauthoryear{{Cooper} et~al.,}{{Cooper}
  et~al.}{2008}]{Cooper2008}
{Cooper} M.~C.,  et~al., 2008, \mn@doi [\mnras]
  {10.1111/j.1365-2966.2007.12613.x}, \href
  {https://ui.adsabs.harvard.edu/abs/2008MNRAS.383.1058C} {383, 1058}

\bibitem[\protect\citeauthoryear{{Correa}, {Schaye}  \& {Trayford}}{{Correa}
  et~al.}{2019}]{Correa2019}
{Correa} C.~A.,  {Schaye} J.,   {Trayford} J.~W.,  2019, \mn@doi [\mnras]
  {10.1093/mnras/stz295}, \href
  {https://ui.adsabs.harvard.edu/abs/2019MNRAS.484.4401C} {484, 4401}

\bibitem[\protect\citeauthoryear{{Croton} et~al.,}{{Croton}
  et~al.}{2006}]{Croton2006}
{Croton} D.~J.,  et~al., 2006, \mn@doi [\mnras]
  {10.1111/j.1365-2966.2005.09675.x}, \href
  {https://ui.adsabs.harvard.edu/abs/2006MNRAS.365...11C} {365, 11}

\bibitem[\protect\citeauthoryear{{Darvish}, {Mobasher}, {Sobral}, {Rettura},
  {Scoville}, {Faisst}  \& {Capak}}{{Darvish} et~al.}{2016}]{Darvish2016}
{Darvish} B.,  {Mobasher} B.,  {Sobral} D.,  {Rettura} A.,  {Scoville} N.,
  {Faisst} A.,   {Capak} P.,  2016, \mn@doi [\apj]
  {10.3847/0004-637X/825/2/113}, \href
  {https://ui.adsabs.harvard.edu/abs/2016ApJ...825..113D} {825, 113}

\bibitem[\protect\citeauthoryear{{Dav{\'e}} et~al.,}{{Dav{\'e}}
  et~al.}{2001}]{Dave2001}
{Dav{\'e}} R.,  et~al., 2001, \mn@doi [\apj] {10.1086/320548}, \href
  {https://ui.adsabs.harvard.edu/abs/2001ApJ...552..473D} {552, 473}

\bibitem[\protect\citeauthoryear{{Dav{\'e}}, {Angl{\'e}s-Alc{\'a}zar},
  {Narayanan}, {Li}, {Rafieferantsoa}  \& {Appleby}}{{Dav{\'e}}
  et~al.}{2019}]{Dave2019}
{Dav{\'e}} R.,  {Angl{\'e}s-Alc{\'a}zar} D.,  {Narayanan} D.,  {Li} Q.,
  {Rafieferantsoa} M.~H.,   {Appleby} S.,  2019, \mn@doi [\mnras]
  {10.1093/mnras/stz937}, \href
  {https://ui.adsabs.harvard.edu/abs/2019MNRAS.486.2827D} {486, 2827}

\bibitem[\protect\citeauthoryear{{Davidzon} et~al.,}{{Davidzon}
  et~al.}{2017}]{Davidzon2017}
{Davidzon} I.,  et~al., 2017, \mn@doi [\aap] {10.1051/0004-6361/201730419},
  \href {https://ui.adsabs.harvard.edu/abs/2017A&A...605A..70D} {605, A70}

\bibitem[\protect\citeauthoryear{{Davies}, {Crain}, {McCarthy}, {Oppenheimer},
  {Schaye}, {Schaller}  \& {McAlpine}}{{Davies} et~al.}{2019}]{Davies2019}
{Davies} J.~J.,  {Crain} R.~A.,  {McCarthy} I.~G.,  {Oppenheimer} B.~D.,
  {Schaye} J.,  {Schaller} M.,   {McAlpine} S.,  2019, \mn@doi [\mnras]
  {10.1093/mnras/stz635}, \href
  {https://ui.adsabs.harvard.edu/abs/2019MNRAS.485.3783D} {485, 3783}

\bibitem[\protect\citeauthoryear{{Davies}, {Crain}, {Oppenheimer}  \&
  {Schaye}}{{Davies} et~al.}{2020}]{Davies2020}
{Davies} J.~J.,  {Crain} R.~A.,  {Oppenheimer} B.~D.,   {Schaye} J.,  2020,
  \mn@doi [\mnras] {10.1093/mnras/stz3201}, \href
  {https://ui.adsabs.harvard.edu/abs/2020MNRAS.491.4462D} {491, 4462}

\bibitem[\protect\citeauthoryear{{Dekel} \& {Birnboim}}{{Dekel} \&
  {Birnboim}}{2006}]{DekelBirnboim2006}
{Dekel} A.,  {Birnboim} Y.,  2006, \mn@doi [\mnras]
  {10.1111/j.1365-2966.2006.10145.x}, \href
  {https://ui.adsabs.harvard.edu/abs/2006MNRAS.368....2D} {368, 2}

\bibitem[\protect\citeauthoryear{{Dekel} \& {Silk}}{{Dekel} \&
  {Silk}}{1986}]{DekelSilk1986}
{Dekel} A.,  {Silk} J.,  1986, \mn@doi [\apj] {10.1086/164050}, \href
  {https://ui.adsabs.harvard.edu/abs/1986ApJ...303...39D} {303, 39}

\bibitem[\protect\citeauthoryear{{Dou} et~al.,}{{Dou} et~al.}{2021a}]{Dou2021}
{Dou} J.,  et~al., 2021a, \mn@doi [\apj] {10.3847/1538-4357/abd17c}, \href
  {https://ui.adsabs.harvard.edu/abs/2021ApJ...907..114D} {907, 114}

\bibitem[\protect\citeauthoryear{{Dou} et~al.,}{{Dou} et~al.}{2021b}]{Dou2021b}
{Dou} J.,  et~al., 2021b, \mn@doi [\apj] {10.3847/1538-4357/abfaf7}, \href
  {https://ui.adsabs.harvard.edu/abs/2021ApJ...915...94D} {915, 94}

\bibitem[\protect\citeauthoryear{{Dressler}}{{Dressler}}{1980}]{Dressler1980}
{Dressler} A.,  1980, \mn@doi [\apj] {10.1086/157753}, \href
  {https://ui.adsabs.harvard.edu/abs/1980ApJ...236..351D} {236, 351}

\bibitem[\protect\citeauthoryear{{Elahi}, {Ca{\~n}as}, {Poulton}, {Tobar},
  {Willis}, {Lagos}, {Power}  \& {Robotham}}{{Elahi} et~al.}{2019}]{Elahi2019}
{Elahi} P.~J.,  {Ca{\~n}as} R.,  {Poulton} R. J.~J.,  {Tobar} R.~J.,  {Willis}
  J.~S.,  {Lagos} C. d.~P.,  {Power} C.,   {Robotham} A. S.~G.,  2019, \mn@doi
  [\pasa] {10.1017/pasa.2019.12}, \href
  {https://ui.adsabs.harvard.edu/abs/2019PASA...36...21E} {36, e021}

\bibitem[\protect\citeauthoryear{{Elbaz} et~al.,}{{Elbaz}
  et~al.}{2007}]{Elbaz2007}
{Elbaz} D.,  et~al., 2007, \mn@doi [\aap] {10.1051/0004-6361:20077525}, \href
  {https://ui.adsabs.harvard.edu/abs/2007A&A...468...33E} {468, 33}

\bibitem[\protect\citeauthoryear{{Ellison} et~al.,}{{Ellison}
  et~al.}{2020}]{Ellison2020}
{Ellison} S.~L.,  et~al., 2020, \mn@doi [\mnras] {10.1093/mnrasl/slz179}, \href
  {https://ui.adsabs.harvard.edu/abs/2020MNRAS.493L..39E} {493, L39}

\bibitem[\protect\citeauthoryear{{Ellison}, {Lin}, {Thorp}, {Pan},
  {S{\'a}nchez}, {Bluck}  \& {Belfiore}}{{Ellison} et~al.}{2021}]{Ellison2021}
{Ellison} S.~L.,  {Lin} L.,  {Thorp} M.~D.,  {Pan} H.-A.,  {S{\'a}nchez} S.~F.,
   {Bluck} A. F.~L.,   {Belfiore} F.,  2021, \mn@doi [\mnras]
  {10.1093/mnrasl/slaa199}, \href
  {https://ui.adsabs.harvard.edu/abs/2021MNRAS.502L...6E} {502, L6}

\bibitem[\protect\citeauthoryear{Elor \& Averbuch-Elor}{Elor \&
  Averbuch-Elor}{2022}]{ElorAverbuchElor2022}
Elor Y.,  Averbuch-Elor H.,  2022, ArXiv, abs/2201.08528

\bibitem[\protect\citeauthoryear{{Fabian}}{{Fabian}}{2012}]{Fabian2012}
{Fabian} A.~C.,  2012, \mn@doi [\araa] {10.1146/annurev-astro-081811-125521},
  \href {https://ui.adsabs.harvard.edu/abs/2012ARA&A..50..455F} {50, 455}

\bibitem[\protect\citeauthoryear{{Fukugita} \& {Peebles}}{{Fukugita} \&
  {Peebles}}{2004}]{FukugitaPeebles2004}
{Fukugita} M.,  {Peebles} P.~J.~E.,  2004, \mn@doi [\apj] {10.1086/425155},
  \href {https://ui.adsabs.harvard.edu/abs/2004ApJ...616..643F} {616, 643}

\bibitem[\protect\citeauthoryear{{Gabor} \& {Dav{\'e}}}{{Gabor} \&
  {Dav{\'e}}}{2012}]{GaborDave2012}
{Gabor} J.~M.,  {Dav{\'e}} R.,  2012, \mn@doi [\mnras]
  {10.1111/j.1365-2966.2012.21640.x}, \href
  {https://ui.adsabs.harvard.edu/abs/2012MNRAS.427.1816G} {427, 1816}

\bibitem[\protect\citeauthoryear{{Gabor} \& {Dav{\'e}}}{{Gabor} \&
  {Dav{\'e}}}{2015}]{GaborDave2015}
{Gabor} J.~M.,  {Dav{\'e}} R.,  2015, \mn@doi [\mnras] {10.1093/mnras/stu2399},
  \href {https://ui.adsabs.harvard.edu/abs/2015MNRAS.447..374G} {447, 374}

\bibitem[\protect\citeauthoryear{{Gallagher} \& {Ostriker}}{{Gallagher} \&
  {Ostriker}}{1972}]{GallagherOstriker1972}
{Gallagher} III J.~S.,  {Ostriker} J.~P.,  1972, \mn@doi [\aj]
  {10.1086/111280}, \href
  {https://ui.adsabs.harvard.edu/abs/1972AJ.....77..288G} {77, 288}

\bibitem[\protect\citeauthoryear{{Gallazzi}, {Bell}, {Zibetti}, {Brinchmann}
  \& {Kelson}}{{Gallazzi} et~al.}{2014}]{Gallazzi2014}
{Gallazzi} A.,  {Bell} E.~F.,  {Zibetti} S.,  {Brinchmann} J.,   {Kelson}
  D.~D.,  2014, \mn@doi [\apj] {10.1088/0004-637X/788/1/72}, \href
  {https://ui.adsabs.harvard.edu/abs/2014ApJ...788...72G} {788, 72}

\bibitem[\protect\citeauthoryear{{Genzel} et~al.,}{{Genzel}
  et~al.}{2015}]{Genzel2015}
{Genzel} R.,  et~al., 2015, \mn@doi [\apj] {10.1088/0004-637X/800/1/20}, \href
  {https://ui.adsabs.harvard.edu/abs/2015ApJ...800...20G} {800, 20}

\bibitem[\protect\citeauthoryear{{Goubert}, {Bluck}, {Piotrowska}  \&
  {Maiolino}}{{Goubert} et~al.}{2024}]{Goubert2024}
{Goubert} P.~H.,  {Bluck} A. F.~L.,  {Piotrowska} J.~M.,   {Maiolino} R.,
  2024, \mn@doi [\mnras] {10.1093/mnras/stae269}, \href
  {https://ui.adsabs.harvard.edu/abs/2024MNRAS.528.4891G} {528, 4891}

\bibitem[\protect\citeauthoryear{{Greene}, {Zakamska}  \& {Smith}}{{Greene}
  et~al.}{2012}]{GreeneZakamskaSmith2012}
{Greene} J.~E.,  {Zakamska} N.~L.,   {Smith} P.~S.,  2012, \mn@doi [\apj]
  {10.1088/0004-637X/746/1/86}, \href
  {https://ui.adsabs.harvard.edu/abs/2012ApJ...746...86G} {746, 86}

\bibitem[\protect\citeauthoryear{{Gunn} \& {Gott}}{{Gunn} \&
  {Gott}}{1972}]{GunnGott1972}
{Gunn} J.~E.,  {Gott} III J.~R.,  1972, \mn@doi [\apj] {10.1086/151605}, \href
  {https://ui.adsabs.harvard.edu/abs/1972ApJ...176....1G} {176, 1}

\bibitem[\protect\citeauthoryear{{Hambrick}, {Ostriker}, {Naab}  \&
  {Johansson}}{{Hambrick} et~al.}{2011}]{Hambrick2011}
{Hambrick} D.~C.,  {Ostriker} J.~P.,  {Naab} T.,   {Johansson} P.~H.,  2011,
  \mn@doi [\apj] {10.1088/0004-637X/738/1/16}, \href
  {https://ui.adsabs.harvard.edu/abs/2011ApJ...738...16H} {738, 16}

\bibitem[\protect\citeauthoryear{{Harikane} et~al.,}{{Harikane}
  et~al.}{2023}]{Harikane2023}
{Harikane} Y.,  et~al., 2023, \mn@doi [\apj] {10.3847/1538-4357/ad029e}, \href
  {https://ui.adsabs.harvard.edu/abs/2023ApJ...959...39H} {959, 39}

\bibitem[\protect\citeauthoryear{{Harrison}, {Alexander}, {Mullaney}  \&
  {Swinbank}}{{Harrison} et~al.}{2014}]{Harrison2014}
{Harrison} C.~M.,  {Alexander} D.~M.,  {Mullaney} J.~R.,   {Swinbank} A.~M.,
  2014, \mn@doi [\mnras] {10.1093/mnras/stu515}, \href
  {https://ui.adsabs.harvard.edu/abs/2014MNRAS.441.3306H} {441, 3306}

\bibitem[\protect\citeauthoryear{{Hasan} et~al.,}{{Hasan}
  et~al.}{2023}]{Hasan2023}
{Hasan} F.,  et~al., 2023, \mn@doi [\apj] {10.3847/1538-4357/acd11c}, \href
  {https://ui.adsabs.harvard.edu/abs/2023ApJ...950..114H} {950, 114}

\bibitem[\protect\citeauthoryear{{Heckman} \& {Best}}{{Heckman} \&
  {Best}}{2014}]{HeckmanBest2014}
{Heckman} T.~M.,  {Best} P.~N.,  2014, \mn@doi [\araa]
  {10.1146/annurev-astro-081913-035722}, \href
  {https://ui.adsabs.harvard.edu/abs/2014ARA&A..52..589H} {52, 589}

\bibitem[\protect\citeauthoryear{{Jiang}, {Helly}, {Cole}  \& {Frenk}}{{Jiang}
  et~al.}{2014}]{Jiang2014}
{Jiang} L.,  {Helly} J.~C.,  {Cole} S.,   {Frenk} C.~S.,  2014, \mn@doi
  [\mnras] {10.1093/mnras/stu390}, \href
  {https://ui.adsabs.harvard.edu/abs/2014MNRAS.440.2115J} {440, 2115}

\bibitem[\protect\citeauthoryear{{Kauffmann}, {White}, {Heckman}, {M{\'e}nard},
  {Brinchmann}, {Charlot}, {Tremonti}  \& {Brinkmann}}{{Kauffmann}
  et~al.}{2004}]{Kauffmann2004}
{Kauffmann} G.,  {White} S. D.~M.,  {Heckman} T.~M.,  {M{\'e}nard} B.,
  {Brinchmann} J.,  {Charlot} S.,  {Tremonti} C.,   {Brinkmann} J.,  2004,
  \mn@doi [\mnras] {10.1111/j.1365-2966.2004.08117.x}, \href
  {https://ui.adsabs.harvard.edu/abs/2004MNRAS.353..713K} {353, 713}

\bibitem[\protect\citeauthoryear{{Kawinwanichakij} et~al.,}{{Kawinwanichakij}
  et~al.}{2017}]{Kawinwanichakij2017}
{Kawinwanichakij} L.,  et~al., 2017, \mn@doi [\apj] {10.3847/1538-4357/aa8b75},
  \href {https://ui.adsabs.harvard.edu/abs/2017ApJ...847..134K} {847, 134}

\bibitem[\protect\citeauthoryear{{King} \& {Pounds}}{{King} \&
  {Pounds}}{2015}]{KingPounds2015}
{King} A.,  {Pounds} K.,  2015, \mn@doi [\araa]
  {10.1146/annurev-astro-082214-122316}, \href
  {https://ui.adsabs.harvard.edu/abs/2015ARA&A..53..115K} {53, 115}

\bibitem[\protect\citeauthoryear{{Kubo} et~al.,}{{Kubo}
  et~al.}{2022}]{Kubo2022}
{Kubo} M.,  et~al., 2022, \mn@doi [\apj] {10.3847/1538-4357/ac7f2d}, \href
  {https://ui.adsabs.harvard.edu/abs/2022ApJ...935...89K} {935, 89}

\bibitem[\protect\citeauthoryear{{Kugel} et~al.,}{{Kugel}
  et~al.}{2023}]{Kugel2023}
{Kugel} R.,  et~al., 2023, \mn@doi [\mnras] {10.1093/mnras/stad2540}, \href
  {https://ui.adsabs.harvard.edu/abs/2023MNRAS.526.6103K} {526, 6103}

\bibitem[\protect\citeauthoryear{{Lagos} et~al.,}{{Lagos}
  et~al.}{2025}]{Lagos2025}
{Lagos} C. d.~P.,  et~al., 2025, \mn@doi [\mnras] {10.1093/mnras/stae2626},
  \href {https://ui.adsabs.harvard.edu/abs/2025MNRAS.536.2324L} {536, 2324}

\bibitem[\protect\citeauthoryear{{Larson}}{{Larson}}{1974}]{Larson1974}
{Larson} R.~B.,  1974, \mn@doi [\mnras] {10.1093/mnras/169.2.229}, \href
  {https://ui.adsabs.harvard.edu/abs/1974MNRAS.169..229L} {169, 229}

\bibitem[\protect\citeauthoryear{{Larson}, {Tinsley}  \& {Caldwell}}{{Larson}
  et~al.}{1980}]{LarsonTinsleyCaldwell1980}
{Larson} R.~B.,  {Tinsley} B.~M.,   {Caldwell} C.~N.,  1980, \mn@doi [\apj]
  {10.1086/157917}, \href
  {https://ui.adsabs.harvard.edu/abs/1980ApJ...237..692L} {237, 692}

\bibitem[\protect\citeauthoryear{{Leja}, {Tacchella}  \& {Conroy}}{{Leja}
  et~al.}{2019}]{Leja2019}
{Leja} J.,  {Tacchella} S.,   {Conroy} C.,  2019, \mn@doi [\apjl]
  {10.3847/2041-8213/ab2f8c}, \href
  {https://ui.adsabs.harvard.edu/abs/2019ApJ...880L...9L} {880, L9}

\bibitem[\protect\citeauthoryear{{Leja} et~al.,}{{Leja}
  et~al.}{2022}]{Leja2022}
{Leja} J.,  et~al., 2022, \mn@doi [\apj] {10.3847/1538-4357/ac887d}, \href
  {https://ui.adsabs.harvard.edu/abs/2022ApJ...936..165L} {936, 165}

\bibitem[\protect\citeauthoryear{{Lim}, {Mo}, {Wang}  \& {Yang}}{{Lim}
  et~al.}{2016}]{Lim2016}
{Lim} S.~H.,  {Mo} H.~J.,  {Wang} H.,   {Yang} X.,  2016, \mn@doi [\mnras]
  {10.1093/mnras/stv2282}, \href
  {https://ui.adsabs.harvard.edu/abs/2016MNRAS.455..499L} {455, 499}

\bibitem[\protect\citeauthoryear{{Lin} et~al.,}{{Lin} et~al.}{2017a}]{LinL2017}
{Lin} L.,  et~al., 2017a, \mn@doi [\apj] {10.3847/1538-4357/aa96ae}, \href
  {https://ui.adsabs.harvard.edu/abs/2017ApJ...851...18L} {851, 18}

\bibitem[\protect\citeauthoryear{{Lin} et~al.,}{{Lin} et~al.}{2017b}]{Lin2017}
{Lin} Y.-T.,  et~al., 2017b, \mn@doi [\apj] {10.3847/1538-4357/aa9bf5}, \href
  {https://ui.adsabs.harvard.edu/abs/2017ApJ...851..139L} {851, 139}

\bibitem[\protect\citeauthoryear{{Lin} et~al.,}{{Lin} et~al.}{2020}]{LinL2020}
{Lin} L.,  et~al., 2020, \mn@doi [\apj] {10.3847/1538-4357/abba3a}, \href
  {https://ui.adsabs.harvard.edu/abs/2020ApJ...903..145L} {903, 145}

\bibitem[\protect\citeauthoryear{{Liu}, {Zakamska}, {Greene}, {Nesvadba}  \&
  {Liu}}{{Liu} et~al.}{2013}]{Liu2013}
{Liu} G.,  {Zakamska} N.~L.,  {Greene} J.~E.,  {Nesvadba} N. P.~H.,   {Liu} X.,
   2013, \mn@doi [\mnras] {10.1093/mnras/stt1755}, \href
  {https://ui.adsabs.harvard.edu/abs/2013MNRAS.436.2576L} {436, 2576}

\bibitem[\protect\citeauthoryear{{Mac Low} \& {Ferrara}}{{Mac Low} \&
  {Ferrara}}{1999}]{MacLowFerrara1999}
{Mac Low} M.-M.,  {Ferrara} A.,  1999, \mn@doi [\apj] {10.1086/306832}, \href
  {https://ui.adsabs.harvard.edu/abs/1999ApJ...513..142M} {513, 142}

\bibitem[\protect\citeauthoryear{{Maiolino} et~al.,}{{Maiolino}
  et~al.}{2012}]{Maiolino2012}
{Maiolino} R.,  et~al., 2012, \mn@doi [\mnras]
  {10.1111/j.1745-3933.2012.01303.x}, \href
  {https://ui.adsabs.harvard.edu/abs/2012MNRAS.425L..66M} {425, L66}

\bibitem[\protect\citeauthoryear{{Maiolino} et~al.,}{{Maiolino}
  et~al.}{2024a}]{Maiolino2024b}
{Maiolino} R.,  et~al., 2024a, \mn@doi [\nat] {10.1038/s41586-024-07052-5},
  \href {https://ui.adsabs.harvard.edu/abs/2024Natur.627...59M} {627, 59}

\bibitem[\protect\citeauthoryear{{Maiolino} et~al.,}{{Maiolino}
  et~al.}{2024b}]{Maiolino2024a}
{Maiolino} R.,  et~al., 2024b, \mn@doi [\aap] {10.1051/0004-6361/202347640},
  \href {https://ui.adsabs.harvard.edu/abs/2024A&A...691A.145M} {691, A145}

\bibitem[\protect\citeauthoryear{{Mart{\'\i}n-Navarro}, {Brodie}, {Romanowsky},
  {Ruiz-Lara}  \& {van de Ven}}{{Mart{\'\i}n-Navarro}
  et~al.}{2018}]{MartinNavarro2018}
{Mart{\'\i}n-Navarro} I.,  {Brodie} J.~P.,  {Romanowsky} A.~J.,  {Ruiz-Lara}
  T.,   {van de Ven} G.,  2018, \mn@doi [\nat] {10.1038/nature24999}, \href
  {https://ui.adsabs.harvard.edu/abs/2018Natur.553..307M} {553, 307}

\bibitem[\protect\citeauthoryear{{Matthee} et~al.,}{{Matthee}
  et~al.}{2024}]{Matthee2024}
{Matthee} J.,  et~al., 2024, \mn@doi [\apj] {10.3847/1538-4357/ad2345}, \href
  {https://ui.adsabs.harvard.edu/abs/2024ApJ...963..129M} {963, 129}

\bibitem[\protect\citeauthoryear{{McAlpine}, {Bower}, {Rosario}, {Crain},
  {Schaye}  \& {Theuns}}{{McAlpine} et~al.}{2018}]{McAlpine2018}
{McAlpine} S.,  {Bower} R.~G.,  {Rosario} D.~J.,  {Crain} R.~A.,  {Schaye} J.,
   {Theuns} T.,  2018, \mn@doi [\mnras] {10.1093/mnras/sty2489}, \href
  {https://ui.adsabs.harvard.edu/abs/2018MNRAS.481.3118M} {481, 3118}

\bibitem[\protect\citeauthoryear{{McCarthy} et~al.,}{{McCarthy}
  et~al.}{2010}]{McCarthy2010}
{McCarthy} I.~G.,  et~al., 2010, \mn@doi [\mnras]
  {10.1111/j.1365-2966.2010.16750.x}, \href
  {https://ui.adsabs.harvard.edu/abs/2010MNRAS.406..822M} {406, 822}

\bibitem[\protect\citeauthoryear{{McLeod}, {McLure}, {Dunlop}, {Cullen},
  {Carnall}  \& {Duncan}}{{McLeod} et~al.}{2021}]{McLeod2021}
{McLeod} D.~J.,  {McLure} R.~J.,  {Dunlop} J.~S.,  {Cullen} F.,  {Carnall}
  A.~C.,   {Duncan} K.,  2021, \mn@doi [\mnras] {10.1093/mnras/stab731}, \href
  {https://ui.adsabs.harvard.edu/abs/2021MNRAS.503.4413M} {503, 4413}

\bibitem[\protect\citeauthoryear{{McNamara} \& {Nulsen}}{{McNamara} \&
  {Nulsen}}{2007}]{McNamaraNulsen2007}
{McNamara} B.~R.,  {Nulsen} P.~E.~J.,  2007, \mn@doi [\araa]
  {10.1146/annurev.astro.45.051806.110625}, \href
  {https://ui.adsabs.harvard.edu/abs/2007ARA&A..45..117M} {45, 117}

\bibitem[\protect\citeauthoryear{{Merritt}}{{Merritt}}{1984}]{Merritt1984}
{Merritt} D.,  1984, \mn@doi [\apj] {10.1086/161590}, \href
  {https://ui.adsabs.harvard.edu/abs/1984ApJ...276...26M} {276, 26}

\bibitem[\protect\citeauthoryear{{Mitchell} \& {Schaye}}{{Mitchell} \&
  {Schaye}}{2022}]{MitchellSchaye2022}
{Mitchell} P.~D.,  {Schaye} J.,  2022, \mn@doi [\mnras]
  {10.1093/mnras/stab3339}, \href
  {https://ui.adsabs.harvard.edu/abs/2022MNRAS.511.2948M} {511, 2948}

\bibitem[\protect\citeauthoryear{{Mo}, {van den Bosch}  \& {White}}{{Mo}
  et~al.}{2010}]{MovdBWhite2010}
{Mo} H.,  {van den Bosch} F.~C.,   {White} S.,  2010, {Galaxy Formation and
  Evolution}

\bibitem[\protect\citeauthoryear{{Moore}, {Katz}, {Lake}, {Dressler}  \&
  {Oemler}}{{Moore} et~al.}{1996}]{Moore1996}
{Moore} B.,  {Katz} N.,  {Lake} G.,  {Dressler} A.,   {Oemler} A.,  1996,
  \mn@doi [\nat] {10.1038/379613a0}, \href
  {https://ui.adsabs.harvard.edu/abs/1996Natur.379..613M} {379, 613}

\bibitem[\protect\citeauthoryear{{Moster}, {Somerville}, {Maulbetsch}, {van den
  Bosch}, {Macci{\`o}}, {Naab}  \& {Oser}}{{Moster} et~al.}{2010}]{Moster2010}
{Moster} B.~P.,  {Somerville} R.~S.,  {Maulbetsch} C.,  {van den Bosch} F.~C.,
  {Macci{\`o}} A.~V.,  {Naab} T.,   {Oser} L.,  2010, \mn@doi [\apj]
  {10.1088/0004-637X/710/2/903}, \href
  {https://ui.adsabs.harvard.edu/abs/2010ApJ...710..903M} {710, 903}

\bibitem[\protect\citeauthoryear{{Naab} \& {Ostriker}}{{Naab} \&
  {Ostriker}}{2017}]{NaabOstriker2017}
{Naab} T.,  {Ostriker} J.~P.,  2017, \mn@doi [\araa]
  {10.1146/annurev-astro-081913-040019}, \href
  {https://ui.adsabs.harvard.edu/abs/2017ARA&A..55...59N} {55, 59}

\bibitem[\protect\citeauthoryear{{Oppenheimer} et~al.,}{{Oppenheimer}
  et~al.}{2020}]{Oppenheimer2020}
{Oppenheimer} B.~D.,  et~al., 2020, \mn@doi [\mnras] {10.1093/mnras/stz3124},
  \href {https://ui.adsabs.harvard.edu/abs/2020MNRAS.491.2939O} {491, 2939}

\bibitem[\protect\citeauthoryear{{Pacifici} et~al.,}{{Pacifici}
  et~al.}{2016}]{Pacifici2016}
{Pacifici} C.,  et~al., 2016, \mn@doi [\apj] {10.3847/0004-637X/832/1/79},
  \href {https://ui.adsabs.harvard.edu/abs/2016ApJ...832...79P} {832, 79}

\bibitem[\protect\citeauthoryear{{Park} et~al.,}{{Park}
  et~al.}{2022}]{Park2022}
{Park} M.,  et~al., 2022, \mn@doi [\mnras] {10.1093/mnras/stac1773}, \href
  {https://ui.adsabs.harvard.edu/abs/2022MNRAS.515..213P} {515, 213}

\bibitem[\protect\citeauthoryear{{Park} et~al.,}{{Park}
  et~al.}{2024}]{Park2024}
{Park} M.,  et~al., 2024, \mn@doi [\apj] {10.3847/1538-4357/ad7e15}, \href
  {https://ui.adsabs.harvard.edu/abs/2024ApJ...976...72P} {976, 72}

\bibitem[\protect\citeauthoryear{{Patel}, {Holden}, {Kelson}, {Illingworth}  \&
  {Franx}}{{Patel} et~al.}{2009}]{Patel2009}
{Patel} S.~G.,  {Holden} B.~P.,  {Kelson} D.~D.,  {Illingworth} G.~D.,
  {Franx} M.,  2009, \mn@doi [\apjl] {10.1088/0004-637X/705/1/L67}, \href
  {https://ui.adsabs.harvard.edu/abs/2009ApJ...705L..67P} {705, L67}

\bibitem[\protect\citeauthoryear{Pedregosa et~al.,}{Pedregosa
  et~al.}{2011}]{scikit-learn}
Pedregosa F.,  et~al., 2011, Journal of Machine Learning Research, 12, 2825

\bibitem[\protect\citeauthoryear{{Peng} et~al.,}{{Peng}
  et~al.}{2010}]{Peng2010}
{Peng} Y.-j.,  et~al., 2010, \mn@doi [\apj] {10.1088/0004-637X/721/1/193},
  \href {https://ui.adsabs.harvard.edu/abs/2010ApJ...721..193P} {721, 193}

\bibitem[\protect\citeauthoryear{{Peng}, {Lilly}, {Renzini}  \&
  {Carollo}}{{Peng} et~al.}{2012}]{Peng2012}
{Peng} Y.-j.,  {Lilly} S.~J.,  {Renzini} A.,   {Carollo} M.,  2012, \mn@doi
  [\apj] {10.1088/0004-637X/757/1/4}, \href
  {https://ui.adsabs.harvard.edu/abs/2012ApJ...757....4P} {757, 4}

\bibitem[\protect\citeauthoryear{{Peng}, {Maiolino}  \& {Cochrane}}{{Peng}
  et~al.}{2015}]{PengMaiolinoCochrane2015}
{Peng} Y.,  {Maiolino} R.,   {Cochrane} R.,  2015, \mn@doi [\nat]
  {10.1038/nature14439}, \href
  {https://ui.adsabs.harvard.edu/abs/2015Natur.521..192P} {521, 192}

\bibitem[\protect\citeauthoryear{{Perna}, {Lanzuisi}, {Brusa}, {Mignoli}  \&
  {Cresci}}{{Perna} et~al.}{2017}]{Perna2017}
{Perna} M.,  {Lanzuisi} G.,  {Brusa} M.,  {Mignoli} M.,   {Cresci} G.,  2017,
  \mn@doi [\aap] {10.1051/0004-6361/201630369}, \href
  {https://ui.adsabs.harvard.edu/abs/2017A&A...603A..99P} {603, A99}

\bibitem[\protect\citeauthoryear{{Pintos-Castro}, {Yee}, {Muzzin}, {Old}  \&
  {Wilson}}{{Pintos-Castro} et~al.}{2019}]{PintosCastro2019}
{Pintos-Castro} I.,  {Yee} H.~K.~C.,  {Muzzin} A.,  {Old} L.,   {Wilson} G.,
  2019, \mn@doi [\apj] {10.3847/1538-4357/ab14ee}, \href
  {https://ui.adsabs.harvard.edu/abs/2019ApJ...876...40P} {876, 40}

\bibitem[\protect\citeauthoryear{{Piotrowska}, {Bluck}, {Maiolino}, {Concas}
  \& {Peng}}{{Piotrowska} et~al.}{2020}]{Piotrowska2020}
{Piotrowska} J.~M.,  {Bluck} A. F.~L.,  {Maiolino} R.,  {Concas} A.,   {Peng}
  Y.,  2020, \mn@doi [\mnras] {10.1093/mnrasl/slz172}, \href
  {https://ui.adsabs.harvard.edu/abs/2020MNRAS.492L...6P} {492, L6}

\bibitem[\protect\citeauthoryear{{Piotrowska}, {Bluck}, {Maiolino}  \&
  {Peng}}{{Piotrowska} et~al.}{2022}]{Piotrowska2022}
{Piotrowska} J.~M.,  {Bluck} A. F.~L.,  {Maiolino} R.,   {Peng} Y.,  2022,
  \mn@doi [\mnras] {10.1093/mnras/stab3673}, \href
  {https://ui.adsabs.harvard.edu/abs/2022MNRAS.512.1052P} {512, 1052}

\bibitem[\protect\citeauthoryear{{Ploeckinger} \& {Schaye}}{{Ploeckinger} \&
  {Schaye}}{2020}]{PloeckingerSchaye2020}
{Ploeckinger} S.,  {Schaye} J.,  2020, \mn@doi [\mnras]
  {10.1093/mnras/staa2172}, \href
  {https://ui.adsabs.harvard.edu/abs/2020MNRAS.497.4857P} {497, 4857}

\bibitem[\protect\citeauthoryear{{Rodr{\'\i}guez Montero}, {Dav{\'e}}, {Wild},
  {Angl{\'e}s-Alc{\'a}zar}  \& {Narayanan}}{{Rodr{\'\i}guez Montero}
  et~al.}{2019}]{RM2019}
{Rodr{\'\i}guez Montero} F.,  {Dav{\'e}} R.,  {Wild} V.,
  {Angl{\'e}s-Alc{\'a}zar} D.,   {Narayanan} D.,  2019, \mn@doi [\mnras]
  {10.1093/mnras/stz2580}, \href
  {https://ui.adsabs.harvard.edu/abs/2019MNRAS.490.2139R} {490, 2139}

\bibitem[\protect\citeauthoryear{{Sazonov}, {Ostriker}  \& {Sunyaev}}{{Sazonov}
  et~al.}{2004}]{SazonovOstrikerSunyaev2004}
{Sazonov} S.~Y.,  {Ostriker} J.~P.,   {Sunyaev} R.~A.,  2004, \mn@doi [\mnras]
  {10.1111/j.1365-2966.2004.07184.x}, \href
  {https://ui.adsabs.harvard.edu/abs/2004MNRAS.347..144S} {347, 144}

\bibitem[\protect\citeauthoryear{{Scannapieco} \& {Oh}}{{Scannapieco} \&
  {Oh}}{2004}]{ScannapiecoOh2004}
{Scannapieco} E.,  {Oh} S.~P.,  2004, \mn@doi [\apj] {10.1086/386542}, \href
  {https://ui.adsabs.harvard.edu/abs/2004ApJ...608...62S} {608, 62}

\bibitem[\protect\citeauthoryear{{Schaller} et~al.,}{{Schaller}
  et~al.}{2024}]{Schaller2024}
{Schaller} M.,  et~al., 2024, \mn@doi [\mnras] {10.1093/mnras/stae922}, \href
  {https://ui.adsabs.harvard.edu/abs/2024MNRAS.530.2378S} {530, 2378}

\bibitem[\protect\citeauthoryear{{Schaye} \& {Dalla Vecchia}}{{Schaye} \&
  {Dalla Vecchia}}{2008}]{SchayeDallaVecchia2008}
{Schaye} J.,  {Dalla Vecchia} C.,  2008, \mn@doi [\mnras]
  {10.1111/j.1365-2966.2007.12639.x}, \href
  {https://ui.adsabs.harvard.edu/abs/2008MNRAS.383.1210S} {383, 1210}

\bibitem[\protect\citeauthoryear{{Schaye} et~al.,}{{Schaye}
  et~al.}{2015}]{Schaye2015}
{Schaye} J.,  et~al., 2015, \mn@doi [\mnras] {10.1093/mnras/stu2058}, \href
  {https://ui.adsabs.harvard.edu/abs/2015MNRAS.446..521S} {446, 521}

\bibitem[\protect\citeauthoryear{{Schaye} et~al.,}{{Schaye}
  et~al.}{2023}]{Schaye2023}
{Schaye} J.,  et~al., 2023, \mn@doi [\mnras] {10.1093/mnras/stad2419}, \href
  {https://ui.adsabs.harvard.edu/abs/2023MNRAS.526.4978S} {526, 4978}

\bibitem[\protect\citeauthoryear{{Scholtz} et~al.,}{{Scholtz}
  et~al.}{2024}]{Scholtz2024}
{Scholtz} J.,  et~al., 2024, \mn@doi [\aap] {10.1051/0004-6361/202347187},
  \href {https://ui.adsabs.harvard.edu/abs/2024A&A...687A.283S} {687, A283}

\bibitem[\protect\citeauthoryear{{Scoville} et~al.,}{{Scoville}
  et~al.}{2013}]{Scoville2013}
{Scoville} N.,  et~al., 2013, \mn@doi [\apjs] {10.1088/0067-0049/206/1/3},
  \href {https://ui.adsabs.harvard.edu/abs/2013ApJS..206....3S} {206, 3}

\bibitem[\protect\citeauthoryear{{Shimakawa} et~al.,}{{Shimakawa}
  et~al.}{2018}]{Shimakawa2018}
{Shimakawa} R.,  et~al., 2018, \mn@doi [\mnras] {10.1093/mnras/stx2494}, \href
  {https://ui.adsabs.harvard.edu/abs/2018MNRAS.473.1977S} {473, 1977}

\bibitem[\protect\citeauthoryear{{Sijacki}, {Springel}, {Di Matteo}  \&
  {Hernquist}}{{Sijacki} et~al.}{2007}]{Sijacki2007}
{Sijacki} D.,  {Springel} V.,  {Di Matteo} T.,   {Hernquist} L.,  2007, \mn@doi
  [\mnras] {10.1111/j.1365-2966.2007.12153.x}, \href
  {https://ui.adsabs.harvard.edu/abs/2007MNRAS.380..877S} {380, 877}

\bibitem[\protect\citeauthoryear{{Silk} \& {Rees}}{{Silk} \&
  {Rees}}{1998}]{SilkRees1998}
{Silk} J.,  {Rees} M.~J.,  1998, \mn@doi [\aap]
  {10.48550/arXiv.astro-ph/9801013}, \href
  {https://ui.adsabs.harvard.edu/abs/1998A&A...331L...1S} {331, L1}

\bibitem[\protect\citeauthoryear{{Somerville} \& {Dav{\'e}}}{{Somerville} \&
  {Dav{\'e}}}{2015}]{SomervilleDave2015}
{Somerville} R.~S.,  {Dav{\'e}} R.,  2015, \mn@doi [\araa]
  {10.1146/annurev-astro-082812-140951}, \href
  {https://ui.adsabs.harvard.edu/abs/2015ARA&A..53...51S} {53, 51}

\bibitem[\protect\citeauthoryear{{Somerville}, {Hopkins}, {Cox}, {Robertson}
  \& {Hernquist}}{{Somerville} et~al.}{2008}]{Somerville2008}
{Somerville} R.~S.,  {Hopkins} P.~F.,  {Cox} T.~J.,  {Robertson} B.~E.,
  {Hernquist} L.,  2008, \mn@doi [\mnras] {10.1111/j.1365-2966.2008.13805.x},
  \href {https://ui.adsabs.harvard.edu/abs/2008MNRAS.391..481S} {391, 481}

\bibitem[\protect\citeauthoryear{{Speagle}, {Steinhardt}, {Capak}  \&
  {Silverman}}{{Speagle} et~al.}{2014}]{Speagle2014}
{Speagle} J.~S.,  {Steinhardt} C.~L.,  {Capak} P.~L.,   {Silverman} J.~D.,
  2014, \mn@doi [\apjs] {10.1088/0067-0049/214/2/15}, \href
  {https://ui.adsabs.harvard.edu/abs/2014ApJS..214...15S} {214, 15}

\bibitem[\protect\citeauthoryear{{Springel} \& {Hernquist}}{{Springel} \&
  {Hernquist}}{2003}]{SpringelHernquist2003}
{Springel} V.,  {Hernquist} L.,  2003, \mn@doi [\mnras]
  {10.1046/j.1365-8711.2003.06206.x}, \href
  {https://ui.adsabs.harvard.edu/abs/2003MNRAS.339..289S} {339, 289}

\bibitem[\protect\citeauthoryear{{Springel}, {Di Matteo}  \&
  {Hernquist}}{{Springel} et~al.}{2005}]{Springel2005}
{Springel} V.,  {Di Matteo} T.,   {Hernquist} L.,  2005, \mn@doi [\mnras]
  {10.1111/j.1365-2966.2005.09238.x}, \href
  {https://ui.adsabs.harvard.edu/abs/2005MNRAS.361..776S} {361, 776}

\bibitem[\protect\citeauthoryear{{Sturm} et~al.,}{{Sturm}
  et~al.}{2011}]{Sturm2011}
{Sturm} E.,  et~al., 2011, \mn@doi [\apjl] {10.1088/2041-8205/733/1/L16}, \href
  {https://ui.adsabs.harvard.edu/abs/2011ApJ...733L..16S} {733, L16}

\bibitem[\protect\citeauthoryear{{Tacchella} et~al.,}{{Tacchella}
  et~al.}{2022}]{Tacchella2022}
{Tacchella} S.,  et~al., 2022, \mn@doi [\apj] {10.3847/1538-4357/ac449b}, \href
  {https://ui.adsabs.harvard.edu/abs/2022ApJ...926..134T} {926, 134}

\bibitem[\protect\citeauthoryear{{Tacchella} et~al.,}{{Tacchella}
  et~al.}{2024}]{Tacchella2024}
{Tacchella} S.,  et~al., 2024, \mn@doi [arXiv e-prints]
  {10.48550/arXiv.2404.02194}, \href
  {https://ui.adsabs.harvard.edu/abs/2024arXiv240402194T} {p. arXiv:2404.02194}

\bibitem[\protect\citeauthoryear{{Tacconi} et~al.,}{{Tacconi}
  et~al.}{2018}]{Tacconi2018}
{Tacconi} L.~J.,  et~al., 2018, \mn@doi [\apj] {10.3847/1538-4357/aaa4b4},
  \href {https://ui.adsabs.harvard.edu/abs/2018ApJ...853..179T} {853, 179}

\bibitem[\protect\citeauthoryear{{Terrazas}, {Bell}, {Henriques}, {White},
  {Cattaneo}  \& {Woo}}{{Terrazas} et~al.}{2016}]{Terrazas2016}
{Terrazas} B.~A.,  {Bell} E.~F.,  {Henriques} B. M.~B.,  {White} S. D.~M.,
  {Cattaneo} A.,   {Woo} J.,  2016, \mn@doi [\apjl]
  {10.3847/2041-8205/830/1/L12}, \href
  {https://ui.adsabs.harvard.edu/abs/2016ApJ...830L..12T} {830, L12}

\bibitem[\protect\citeauthoryear{{Terrazas}, {Bell}, {Woo}  \&
  {Henriques}}{{Terrazas} et~al.}{2017}]{Terrazas2017}
{Terrazas} B.~A.,  {Bell} E.~F.,  {Woo} J.,   {Henriques} B. M.~B.,  2017,
  \mn@doi [\apj] {10.3847/1538-4357/aa7d07}, \href
  {https://ui.adsabs.harvard.edu/abs/2017ApJ...844..170T} {844, 170}

\bibitem[\protect\citeauthoryear{{Terrazas} et~al.,}{{Terrazas}
  et~al.}{2020}]{Terrazas2020}
{Terrazas} B.~A.,  et~al., 2020, \mn@doi [\mnras] {10.1093/mnras/staa374},
  \href {https://ui.adsabs.harvard.edu/abs/2020MNRAS.493.1888T} {493, 1888}

\bibitem[\protect\citeauthoryear{{Tomczak} et~al.,}{{Tomczak}
  et~al.}{2016}]{Tomczak2016}
{Tomczak} A.~R.,  et~al., 2016, \mn@doi [\apj] {10.3847/0004-637X/817/2/118},
  \href {https://ui.adsabs.harvard.edu/abs/2016ApJ...817..118T} {817, 118}

\bibitem[\protect\citeauthoryear{{Trayford} et~al.,}{{Trayford}
  et~al.}{2015}]{Trayford2015}
{Trayford} J.~W.,  et~al., 2015, \mn@doi [\mnras] {10.1093/mnras/stv1461},
  \href {https://ui.adsabs.harvard.edu/abs/2015MNRAS.452.2879T} {452, 2879}

\bibitem[\protect\citeauthoryear{{Trayford}, {Theuns}, {Bower}, {Crain},
  {Lagos}, {Schaller}  \& {Schaye}}{{Trayford} et~al.}{2016}]{Trayford2016}
{Trayford} J.~W.,  {Theuns} T.,  {Bower} R.~G.,  {Crain} R.~A.,  {Lagos} C.
  d.~P.,  {Schaller} M.,   {Schaye} J.,  2016, \mn@doi [\mnras]
  {10.1093/mnras/stw1230}, \href
  {https://ui.adsabs.harvard.edu/abs/2016MNRAS.460.3925T} {460, 3925}

\bibitem[\protect\citeauthoryear{{{\"U}bler} et~al.,}{{{\"U}bler}
  et~al.}{2023}]{Ubler2023}
{{\"U}bler} H.,  et~al., 2023, \mn@doi [\aap] {10.1051/0004-6361/202346137},
  \href {https://ui.adsabs.harvard.edu/abs/2023A&A...677A.145U} {677, A145}

\bibitem[\protect\citeauthoryear{{Valentino} et~al.,}{{Valentino}
  et~al.}{2023}]{Valentino2023}
{Valentino} F.,  et~al., 2023, \mn@doi [\apj] {10.3847/1538-4357/acbefa}, \href
  {https://ui.adsabs.harvard.edu/abs/2023ApJ...947...20V} {947, 20}

\bibitem[\protect\citeauthoryear{{Veilleux}, {Cecil}  \&
  {Bland-Hawthorn}}{{Veilleux} et~al.}{2005}]{Veilleux2005}
{Veilleux} S.,  {Cecil} G.,   {Bland-Hawthorn} J.,  2005, \mn@doi [\araa]
  {10.1146/annurev.astro.43.072103.150610}, \href
  {https://ui.adsabs.harvard.edu/abs/2005ARA&A..43..769V} {43, 769}

\bibitem[\protect\citeauthoryear{{Vogelsberger} et~al.,}{{Vogelsberger}
  et~al.}{2014}]{Vogelsberger2014}
{Vogelsberger} M.,  et~al., 2014, \mn@doi [\mnras] {10.1093/mnras/stu1536},
  \href {https://ui.adsabs.harvard.edu/abs/2014MNRAS.444.1518V} {444, 1518}

\bibitem[\protect\citeauthoryear{{Weaver} et~al.,}{{Weaver}
  et~al.}{2023}]{Weaver2023}
{Weaver} J.~R.,  et~al., 2023, \mn@doi [\aap] {10.1051/0004-6361/202245581},
  \href {https://ui.adsabs.harvard.edu/abs/2023A&A...677A.184W} {677, A184}

\bibitem[\protect\citeauthoryear{{Werner}, {McNamara}, {Churazov}  \&
  {Scannapieco}}{{Werner} et~al.}{2019}]{Werner2019}
{Werner} N.,  {McNamara} B.~R.,  {Churazov} E.,   {Scannapieco} E.,  2019,
  \mn@doi [\ssr] {10.1007/s11214-018-0571-9}, \href
  {https://ui.adsabs.harvard.edu/abs/2019SSRv..215....5W} {215, 5}

\bibitem[\protect\citeauthoryear{{White} \& {Frenk}}{{White} \&
  {Frenk}}{1991}]{WhiteFrenk1991}
{White} S. D.~M.,  {Frenk} C.~S.,  1991, \mn@doi [\apj] {10.1086/170483}, \href
  {https://ui.adsabs.harvard.edu/abs/1991ApJ...379...52W} {379, 52}

\bibitem[\protect\citeauthoryear{{White} \& {Rees}}{{White} \&
  {Rees}}{1978}]{WhiteRees1978}
{White} S.~D.~M.,  {Rees} M.~J.,  1978, \mn@doi [\mnras]
  {10.1093/mnras/183.3.341}, \href
  {https://ui.adsabs.harvard.edu/abs/1978MNRAS.183..341W} {183, 341}

\bibitem[\protect\citeauthoryear{{Wild} et~al.,}{{Wild}
  et~al.}{2020}]{Wild2020}
{Wild} V.,  et~al., 2020, \mn@doi [\mnras] {10.1093/mnras/staa674}, \href
  {https://ui.adsabs.harvard.edu/abs/2020MNRAS.494..529W} {494, 529}

\bibitem[\protect\citeauthoryear{{Wu} et~al.,}{{Wu} et~al.}{2018}]{Wu2018}
{Wu} P.-F.,  et~al., 2018, \mn@doi [\apj] {10.3847/1538-4357/aae822}, \href
  {https://ui.adsabs.harvard.edu/abs/2018ApJ...868...37W} {868, 37}

\bibitem[\protect\citeauthoryear{{Zinger} et~al.,}{{Zinger}
  et~al.}{2020}]{Zinger2020}
{Zinger} E.,  et~al., 2020, \mn@doi [\mnras] {10.1093/mnras/staa2607}, \href
  {https://ui.adsabs.harvard.edu/abs/2020MNRAS.499..768Z} {499, 768}

\makeatother
\end{thebibliography}



\label{lastpage}

\end{document}